\documentclass[
aps,
prb,
twocolumn,
groupedaddress,
amsmath,amssymb,
]{revtex4-1}

\usepackage{graphicx}
\usepackage{dcolumn}
\usepackage{bm}

\begin{document}

\preprint{APS/123-QED}

\title{A Multi-scale Approach for Simulations of Kelvin
Probe Force Microscopy with Atomic Resolution}


\author{Ali Sadeghi, Alexis Baratoff, S. Alireza Ghasemi, Stefan Goedecker, Thilo Glatzel, Shigeki Kawai and Ernst Meyer}
\affiliation{Department of Physics, University of Basel,
Klingelbergstrasse 82, CH-4056 Basel, Switzerland}




\date{\today}
\begin{abstract}
The distance dependence and atomic-scale contrast recently
observed in nominal contact potential difference (CPD) signals
simultaneously recorded by KPFM using non-contact atomic force microscopy
(NCAFM) on defect-free surfaces of insulating, as well as
semiconducting samples, have stimulated theoretical attempts to
explain such effects. Especially in the case of insulators, it is
not quite clear how the applied bias voltage affects electrostatic
forces acting on the atomic scale. We attack this problem in two
steps. First, the electrostatics of the macroscopic
tip-cantilever-sample system is treated by a finite-difference
method on an adjustable nonuniform mesh. Then the resulting
electric field under the tip apex is inserted into a series of
atomistic wavelet-based density functional theory (DFT)
calculations. Results are shown for a realistic neutral but
reactive silicon nano-scale tip interacting with a NaCl(001)
sample. Bias-dependent forces and resulting atomic displacements
are computed to within an unprecedented accuracy.
 Theoretical
expressions for amplitude modulation (AM) and frequency modulation
(FM) KPFM signals and for the corresponding local contact
potential differences (LCPD) are obtained by combining the
macroscopic and atomistic contributions to the electrostatic force
component generated at the voltage modulation frequency, and
evaluated for several tip oscillation amplitudes $A$ up to 10~nm.
For $A$ = 0.1~\AA, the computed LCPD contrast is proportional to the
slope of the atomistic force versus bias in the AM mode and to its
derivative with respect to the tip-sample separation in the FM
mode.  Being essentially constant over a few Volts, this slope is
the basic quantity which determines variations of the atomic-scale
LCPD contrast.
Already above $A$ = 1~\AA, the LCPD contrasts in both modes
exhibit almost the same spatial dependence as the slope. In the AM
mode, this contrast is approximately proportional to $A^{-1/2}$,
but remains much weaker than the contrast in the FM mode, which
drops somewhat faster as $A$ is increased. These trends are a
consequence of the macroscopic contributions to the KPFM signal,
which are stronger in the AM-mode and
especially important if the sample is an insulator
even at sub-nanometer separations where atomic-scale contrast
appears.

\begin{description}
\item[PACS numbers] {68.37.Ps, 07.79.Lh, 02.70.Bf, 31.15.A-}
\end{description}
\end{abstract}

\maketitle

\section{\label{sec:Introd}Introduction}
\emph{Kelvin probe force microscopy} (KPFM), which was introduced
twenty years ago,\cite{Weaver91,Nonnenmacher91} has become an
attractive non-contact technique to determine the electric surface
characteristics of materials. Among many applications, this
technique has been successfully applied for mapping local work function
or surface potential variations along inhomogeneous surfaces of
various materials.\cite{Sadewasser02,Barth11,KPFMbook} For a conducting
crystal, the work function corresponds to the energy difference
between the vacuum level outside the surface at a distance large
compared to the lattice spacing, yet small compared to the lateral
dimensions of a homogeneous patch, and the bulk Fermi level. In
this range, which is typical for conventional KPFM measurements,
the potential acting on an electron approaches the local vacuum
level and becomes constant, except in the vicinity of surface
steps or patch boundaries. Differences between local vacuum levels
are solely due to electrostatic contributions which give rise to
fringing electric fields around such boundaries.
 If the sample is covered by a thin
overlayer of foreign material, the work function can change owing
to electron transfer and structural relaxation at the
interface.~\cite{Olsson03}  Similar changes can occur at the
surface of a doped semiconductor, owing to band bending in a
subsurface depletion layer.
 As long as electrochemical equilibrium occurs
the Fermi level is aligned throughout the sample and with the
Fermi level of the back-electrode. However, if the sample is a
wide-bandgap insulator, e.g. an alkali halide crystal, this
equilibration may require very long times, so that the bulk Fermi
level is not well-defined.  Charge rearrangements and relaxation
occur at the interface with the back electrode
 and cause an additive shift of the local vacuum level outside the
surface
with respect to the Fermi level of the back electrode. In a real,
thick enough insulator with charged impurities, such a shift will
also be affected by the distribution of spatially separated
charged defects at the interface, the surface and in the bulk of
the sample, as well as by their
slow diffusion over time~\cite{Barth07,Barth11}.

When two separated conducting bodies, e.g. the probe tip of an
\emph{Atomic Force Microscope} (AFM) and the sample, with
different work functions $\phi_{t}$ and $\phi_{s}$ are connected
via back electrodes, electrons are transferred until the Fermi
levels become aligned. The charged bodies then develop a
\emph{contact potential difference} (CPD), of $V_{CPD}= (\phi_{t}
- \phi_{s})/e$, where $e$ is the elementary charge and 
the sample is grounded.
If the tip is biased at $V_{b}$ with respect to the sample, a finite
electric field ${\bf E}\propto V$ develops in the gap between them
and causes an attractive electrostatic force proportional to $V^2$
where $V=V_b-V_{CPD}$ is their \emph{effective potential
difference}.
 If the sample is an insulator the same phenomenon occurs, but
$\phi_{s}$ must be referred to the Fermi level of the
back-electrode and is therefore affected by all the
above-mentioned shifts, and so is $V_{CPD}$.  It is then more
appropriate to focus attention on variations of $V_{CPD}$ along
the surface rather than on its absolute value which is affected by
sample preparation.

In KPFM, a signal determined by this electrostatic force is
compensated by applying a DC bias $V_b=V_{CPD}$. For fast
measurements the applied bias consists of an AC modulation voltage
with angular frequency $\omega = 2 \pi f$ in addition to the DC
voltage:
\begin{eqnarray}
V_b(t)=V_{DC} + V_{AC} \cos \omega t.
 \label{Vt}
\end{eqnarray}
Assuming that the electric response is linear and in-phase with
$V_{AC}$, the electrostatic force acting on the tip can be
decomposed into three spectral components:
\begin{equation}
F_{}(t)=F_{DC}+F_{\omega}\cos \omega t +F_{2\omega}\cos 2\omega t.
\end{equation}
The $\omega$ component of the KPFM signal, which is proportional
to $\left(V_{DC} - V_{CPD}\right) V_{AC}$, is selectively detected
by a lock-in amplifier and compensated by a feedback circuit.

CPD variations along a surface can be conveniently measured
together with its topography,\cite{Nonnenmacher91} as determined
by \emph{non-contact atomic force microscopy} (NCAFM). In most
state-of-the-art NCAFM experiments a micro-fabricated cantilever
with a tip at its free end (typically etched out of doped
single-crystal silicon) oscillates with a constant amplitude $A$
at the frequency $f_1$ of a flexural resonance (usually the
fundamental mode).~\cite{Giessibl95,Giessibl03} 
Distance-dependent tip-sample forces cause a frequency shift
$\Delta f_1$ which can be very accurately measured using FM
detection (frequency demodulation)~\cite{Albrecht91} and used for
distance control. In \emph{combined NCAFM-KPFM}, the $F_{\omega}$
component is simultaneously sensed; either the modulated
deflection signal (\emph{Amplitude Modulation} KPFM~\cite{kikukawa96}) or the
modulation of the resonance frequency shift $\Delta f_1$
(\emph{Frequency Modulation} KPFM~\cite{Kitamura98}) is actually
detected.~\cite{Zerweck05,Giessibl98}  In either case the amplitude of the
signal at the modulation frequency $f$ is proportional to $(V_{DC}
- V_{CPD}) V_{AC}$. Thus $V_{CPD}$ can be recorded by continuously
adjusting $V_{DC}$ so that the modulation signal vanishes while
scanning the tip parallel to the sample surface at a distance
controlled by the (non-modulated) shift $\Delta
f_1$.~\cite{Giessibl03} Both modulation techniques are much faster
and more sensitive than the direct method in which $V_{CPD}$ is
determined from the extremum of the parabolic $\Delta f_1(V_b)$
curve measured by slowly sweeping $V_b$ at each measurement
point.\cite{Guggisberg00,Sadewasser09,Koenig09,Gross09a} Potential
artifacts of the modulation techniques~\cite{Diesinger08}
are avoided in the direct quasistatic method.
 Because the scope of this article is
primarily theoretical, we don't further consider such experimental
difficulties, but focus our attention on still controversial
\emph{atomic-scale variations} of the so-called \emph{local} CPD
or $V_{LCPD}$ on large defect-free surface areas. Thus we
deliberately leave out local changes due to charged surface
defects\cite{Sommerhalter99,Koenig09,Barth11} or
adsorbates\cite{Gross09a,Gross09b} which have recently attracted
considerable attention, also in theory.~\cite{Barth10,Bocquet11}

Atomic-scale variations of $\Delta f_1$ can be detected by
NCAFM on well-prepared surfaces in ultrahigh vacuum if the closest
approach distance of the tip is smaller than the lattice spacing
or the spacing between protruding atoms.\cite{Giessibl95} The
contrast in $\Delta f_1$ then arises from short-range interatomic
forces which begin to act in that distance range, while cantilever
jump-to-contact is avoided if the total force remains much smaller
than the maximum restoring force $k A$, $k$ and $A$ being
respectively the flexural lever stiffness and oscillation
amplitude.\cite{Giessibl03} Combined NCAFM-KPFM experiments have
proven that
FM-KPFM,~\cite{Kitamura00,Okamoto02,Okamoto03b,Krok08,Sadewasser09}
as well as AM-KPFM~\cite{Enevoldsen08,Bocquet08,Kawai10} could
detect lateral atomic-scale variations of $V_{LCPD}$ in the range
where $\Delta f_1$ exhibits similar variations on surfaces of
semiconductors, as well as of ionic crystals.  Understanding the
connections between the observed contrast in $V_{LCPD}$ and the
atomic-scale variations of the electrostatic potential just
outside the surface has been a challenging task, especially on
unreconstructed cleavage faces of rocksalt-type
crystals.~\cite{Bocquet08}
Above a flat homogeneous surface $V_{LCPD}$ must, in principle,
approach the corresponding $V_{CPD}$ at somewhat larger tip-sample
separations.  In practice, however, this ideal behavior is
often masked by a slow dependence caused by the finite lateral
resolution of surface inhomogeneities, e.g. finite islands of
materials with different work functions.  This effect is less
pronounced in FM- than in
AM-KPFM.~\cite{Glatzel03,Zerweck05,Krok08,Glatzel09} Several
researchers developed models and computational schemes based on
classical electrostatics which treated the tip and the sample
(sometimes also the cantilever) as macroscopic bodies in order to
interpret the resolution of KPFM images of inhomogeneous surfaces
on lateral scales of several nanometers and
above.\cite{Hochwitz96,Jacobs98,Belaidi98,Colchero01,Gomez01a,Strassburg05,Konior07,Shen08b}
On the other hand, only few authors considered atomistic
nano-scale tip-sample systems, either
neglecting~\cite{Sadewasser09,Masago10} or including the macroscopic
contributions via simple approximations.
 In the first theoretical study of combined NCAFM-KPFM on an ionic
crystal sample,~\cite{Bocquet08,Nony09a,KPFMbook} a formally correct
partitioning was proposed between capacitive and
short-range electrostatic forces induced by the effective
macroscopic bias $V$.  This analytic treatment also provided
qualitative insights into the origin of atomic-scale $LCPD$ contrast,
although underestimating the capacitive forces caused a quantitatively 
disagreement with experimental results
as will be explained in subsection~\ref{sec:previous}.
More reliable results were obtained 
 for a NaCl(001) sample interacting with a model tip consisting of a conducting sphere terminated by a NaCl cluster
by allowing local atomic deformations.~\cite{Nony09b} These
atomistic simulations were based on the SCIFI
code~\cite{Kantorovich00} which has provided detailed insights
into NCAFM on ionic compounds.~\cite{Hoffmann04,Ruschmeier08}

In the present work, which is based on separate classical
electrostatics and \emph{ab initio} calculations, we propose a more
rigorous and accurate approach for coupling interactions acting on
widely different length scales which leads to an unambiguous
definition of $V_{LCPD}$. The outline of this paper is as follows:
in Section~\ref{sec:macro} we discuss previous approaches, then
present our own computationally simple, yet flexible
finite-difference (FD) scheme with controlled accuracy to treat
electrostatic tip-sample interactions on macro- and mesoscopic
scales. Owing to electric field penetration into the dielectric
sample, the tip shank and the cantilever significantly affect the
capacitive force and its gradient even at sub-nanometer
tip-surface separations where atomic-scale contrast appears. We
also explain how the influence of the effective bias $V$ can be
included into atomistic calculations, as well as shortcomings of
previous attempts to do so.  In Section~\ref{sec:micro} we
critically discuss previous atomistic calculations, as well as
experimental evidence for short-range electrostatic interactions.
Density functional calculations for nano-scale
tip-sample systems are then discussed and illustrated for a
realistic Si tip close to a NaCl(001) slab as an example of
current interest. One important result is that the microscopic
short-range force is proportional to $V$ over a few volts; the
corresponding slope is thus the basic quantity that should be
extracted from KPFM measurements.  In Section~\ref{sec:kpfm}
expressions for $V_{LCPD}$ in AM- and FM-KPFM are obtained and
evaluated, first for ultrasmall, then for finite tip oscillation
amplitude $A$.  Their magnitude and dependence on $A$ are
explained in detail in terms of the above-mentioned macroscopic
contributions to the capacitive force. Experimental limitations
and evidence for the predicted trends, as well as desirable
measurements are also briefly discussed.
Finally, in Section~\ref{sec:summary} the main features of our
approach and of our results are summarized, and conclusions are
drawn.

\section{Macroscopic electrostatic interaction\label{sec:macro}}
\subsection{Previous approaches\label{sec:previous}}
Calculating the cantilever-tip-sample electrostatic interaction is, in fact, 
an intricate electrostatic boundary-value problem. The
main difficulty is due to the presence of several length scales
determined by the nontrivial shape of AFM components, as well as
to the distance-dependent redistribution of the surface charge
density at constant bias voltage. In the case of conductive bodies
with cylindrical symmetry, a simple assumption (uniform electric
field along field lines approximated by circular arcs to their
surfaces) led to an analytic expression for the force on the tip
treated as cone with a spherical end cap.~\cite{Hudlet98}
 Recent numerical calculations
~\cite{Shen08a,Barth10} showed that Hudlet's
expression is surprisingly accurate.  Somewhat different
analytical expressions
 and estimates for the lateral resolution in AM- and FM-KPFM
were obtained
 for similar probes, also including a tilted
cantilever.~\cite{Colchero01}.
 For cylindrical geometries, many authors proposed numerical
schemes based on the image charge method
 which is applicable to simple geometries involving spherical and planar
surfaces.~\cite{Jackson} Thus Belaidi \emph{et al}\cite{Belaidi97}
placed N point charges on the symmetry axis and determined their
positions and strengths by forcing the potential on the tip
surface to be $V$ by a nonlinear least squares fit. The previously
mentioned authors also described how contributions of the
spherical cap, the tip shank and the cantilever to the macroscopic
force lead to characteristic distance dependencies on scales
determined by the geometry and dimensions of those parts.
A \emph{linearized} version of the
\emph{numerical image charge method} where the positions of axial
point and line charges were fixed 
was
applied to study tip-shape effects
for conductive and \emph{dielectric
samples}~\cite{Gomez01a,Gomez01b} \emph{and thin films on
conducting substrates}~\cite{Sacha07}, also including the
influence of the cantilever\cite{Sacha04}.
It is not known to what extent the boundary conditions must be
satisfied for a given accuracy in the numerical image method,
unlike in the analytic method where the positions and strengths of
the image charges change with tip-sample separation 
and the boundary conditions are fully satisfied
(see Appendix~\ref{app:sphere}).

A more systematic approach to
 multi-length-scale problems
 is the \emph{boundary element method}
(BEM)~\cite{Strassburg05,Konior07,Shen08a}. In this method the 3D
(2D) differntial Poisson's equation
 is transformed into 2D (1D) integral (Green's functions) equations
on the surfaces of conductive or dielectric components, including
CPD discontinuities and surface charges if desired.~\cite{Shen08b}
The accuracy of BEM is controlled by the mesh resolution and is
applicable to complex probe-sample systems, e.g. including a
realistic cantilever~\cite{Elias11}.
The size of the resulting linear system of equations is small
compared to volumetric discretization methods.
 However, because of the memory requirement of $\mathcal O(N^2)$ to
store the fully populated matrix and complexity of $\mathcal
O(N^3)$ to solve the linear equations, BEM
has mostly been applied to systems with a relatively small number
$N$ of grid points, e.g.
problems of high symmetry and homogeneity for which it is feasible
to derive the Green's function analytically. Somewhat earlier a
few authors adapted Green's function methods developed for more
complex near-field optics problems to investigate lateral
resolution in KPFM on inhomogeneous
samples~\cite{Jacobs98,Gomez01a}.
One advantage of BEM is that the LCPD of such samples can be
expressed as a 2D convolution of the CPD and/or of a fixed surface
charge distribution with a point-spread function which depends
only on the relative position of the scanning
probe.~\cite{Strassburg05,Shen08a,Elias11}  The distance-dependent
lateral resolution can be quantified by the width of that
function. Moreover, if one assumes that only one of those
distribution is present, its can be determined by inversion of the
BEM matrix upon discretization on the adjustable BEM
mesh.~\cite{Shen08b}

Conceptually more straightforward approaches involving surface
elements have been applied to conductive probe and sample systems.
In the simplest one, the tip surface is approximated as a regular
staircase (or, equivalently, as an \emph{array of capacitors in
parallel}),~\cite{Jacobs98,Sadewasser03,Zerweck05}.  More accurate
methods rely on adjustable meshes. Thus the \emph{finite element
method} (FEM) was used to calculate the electrostatic force acting
on a conical tip,~\cite{Belaidi98}
while a commercial FEM software was recently applied
to simulate a realistic cantilever and tip of actual shape and
dimensions over a \emph {conducting} flat sample with a CPD
discontinuity.~\cite{Valdre08}  More sophisticated software
packages have been used to solve the Poisson's equation in the
presence of space charges, e.g. for structured samples involving
doped semiconductors~\cite{Hochwitz96,Charrier08}. Numerical
methods which involve 3D discretization
require a very large number of grid points even if the mesh is
carefully adjusted;
 the computational box must therefore be truncated at some finite
extent.

\subsection{\label{sec:FDM} Implementation of finite-difference method}
As an alternative
 we present a finite-difference
method (FDM) on a 3D non-uniform grid which is capable of dealing
with realistic sizes of the cantilever, tip and sample.
Inhomogeneous metallic and dielectric samples as well as thin
dielectric films on metal substrates, can be straightforwardly
treated with this method. The most attractive feature of our FDM
compared to FEM or BEM computations is its ease of implementation.
Since the electrostatic potential varies smoothly and slowly at
distances far from the tip apex, we use a grid spacing which
increases exponentially away from this region. Consequently, the
number of grid points depends logarithmically on the truncation
lengths, and an extension of the computational box costs
relatively few additional grid points. It allows us to simulate
the cantilever as well as thick dielectric samples according to
their actual sizes in experiments.

\begin{figure}
\begin{center}
\includegraphics[width=.5\textwidth]{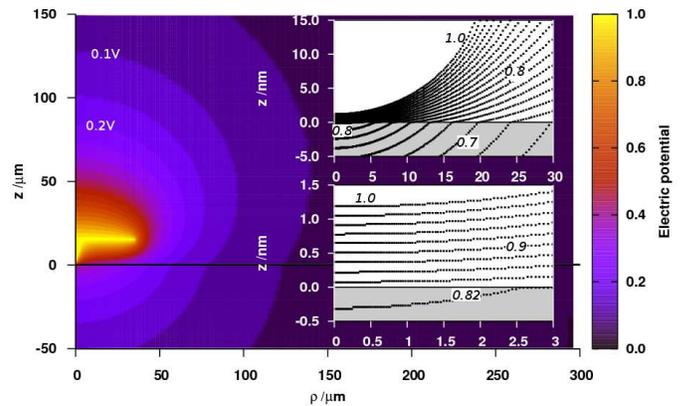}
\caption {\label{fig:map} (color online) 2D ($\rho, z$) maps of
the macroscopic electrostatic potential at three magnifications
($\times10^4$ and $\times10^5$ in zoom-in windows). An effective
bias $V=1$ Volt is applied to the conducting probe while the
back-electrode as well as the surrounding enclosure are grounded.
The yellow region corresponding to $\Phi=1V$ reflects the assumed
cylindrically symmetric probe geometry: a cone with 15$^o$
half-angle terminated by a spherical cap of radius $R$ = 20 nm.
The cone is 15nm high and attached to a disk of thickness 0.5
$\mu$m. The radius of the disk is 35$\mu$m  which matches the area
of a typical cantilever. The sample is a 1-mm thick dielectric
slab with the relative permittivity $\epsilon/\epsilon_0 = 5.9$ of
NaCl. The back electrode and the surrounding enclosure of height
and radius $10^6 R=20$ mm are grounded (not shown). The
sample-vacuum interface is indicated by the horizontal lines at
$z=0$. }
 \end{center}
 \end{figure}

The capacitance $C(s)$ between the probe and the sample
back-electrode depends only on the tip-sample separation $s$,
provided that their geometries are fixed.~\cite{Smythe} The
macroscopic electrostatic energy due to the effective voltage
difference $V=V_b-V_{CPD}$ between the conducting tip and
back-electrode is given by $U_c(s,V)= \frac{1}{2} C(s)V^2$. The
electrostatic force exerted on the tip
is proportional to the capacitance-gradient $C'(s)$ 
\begin{equation}\label{eqn:FM}
F_M(s,V)=-\frac{\partial U}{\partial s} =+\left(\frac{\partial U_c}{\partial s} \right)_{V}=
+\frac{1}{2} C'(s) V^2.
\end{equation}
Similarly, the force-gradient is proportional to $C''(s)=
\partial^2C/\partial s^2$. We wish to emphasize the difference between
$U$ and $U_c$ which leads to the positive sign on the RHS of
Eq.~(\ref{eqn:FM}); the reason is restated for convenience in
Appendix~\ref{app:sign}. The electrostatic energy
\begin{align*}
U_c(s,V)=\frac{1}{2} \int \epsilon({\bf r}) |{\bf \nabla} \Phi |^2 \, d{\bf r}
\end{align*}
can be determined once the electrostatic potential $\Phi({\bf r};s,V)$ is known
at any point $\bf r$ in space.
In general, when the dielectric constant $\epsilon({\bf r})$ varies in space,
$\Phi$ satisfies the generalized form of Poisson's equation 
\begin{eqnarray} \label{eqn:poisson}
{\bf \nabla}  \cdot \left[\epsilon({\bf r})\nabla \Phi ({\bf r})\right]=-\rho({\bf r}),
\end{eqnarray}
$\rho$ being the charge density.
Minimization of the energy-like functional
\begin{align}\label{eqn:I}
I \left[\Psi({\bf r})\right]=\frac{1}{2} \int \epsilon({\bf r}) \left|
\nabla \Psi \right|^2 d{\bf r}  - \int \rho \Psi  d{\bf r}.
\end{align}
subject to Dirichlet boundary conditions leads to $\Phi$, the
solution of the Poisson's equation Eq.(\ref{eqn:poisson}) with the
same boundary conditions.~\cite{Jackson} Using a discretized
variational approach, we therefore minimize the functional
\begin{eqnarray}\label{eqn:DI}
I\left( \{\Psi_{\bf n}\} \right) =\sum_{\bf n} \left( \frac{1}{2}
\epsilon_{\bf n} \left| \nabla \Psi \right|_{\bf n}^2 - \rho_{\bf n}
\Psi_{\bf n} \right) v_{\bf n}
.\end{eqnarray}
On a non-uniform grid, we delimit the volume $v_{\bf n}$ of the
volume element assigned to node $\bf n$ by neighboring nodes.
Then, $\Psi_{\bf n}$, $\rho_{\bf n}$, $\epsilon_{\bf n}$ and the
electric field $-\nabla \Psi_{\bf n}$ are \emph{evaluated at the
center of the volume element} by linear interpolation between the
nodes adjacent to $\bf n$ in orthogonal directions. This ensures
that the field is effectively evaluated to second order in the
product of grid spacings and that discontinuities in $\nabla
\Psi_{\bf n}$ and $\epsilon_{\bf n}$ at material interfaces are
correctly treated. Although the formalism is general and can be
applied to any 3D system on a judiciously chosen nonuniform 3D
orthogonal grid, in the following examples we \emph{consider a
cylindrically symmetric setup without free charges} in order to
 allow comparison with most previous computations. In cylindrical coordinates, each volume element is a truncated
tube of height $h^{(z)}_{k}$ with inner and outer radii
$\rho_{i}$, $\rho_{i+1}$, respectively, and
$v_{\bf n}=\pi(\rho_{i+1} + \rho_i)h^{(\rho)}_{i}h^{(z)}_{k}$,
$h^{(\rho)}_{i}=\rho_{i+1}-\rho_{i}$ and $h^{(z)}_{k}=z_{k+1}-z_k$
being respectively the radial and vertical spacings of the
appropriate nonuniform grid.
 The radial and vertical components of $\nabla \Psi$ are
approximated on the circle of radius $\rho_{i}+0.5h^{(\rho)}_i$ at
$z_k+0.5h^{(z)}_{k}$
 as $ ({\Psi_{i+1,k}-\Psi_{i,k}})/{h^{(\rho)}_{i}}$ and
$({\Psi_{i,k+1}-\Psi_{i,k}})/h^{(z)}_{k}$.
  Since the FD approximation of the electric field
is a linear combination of the potential values on nearest neighbor nodes,
the functional in Eq. (\ref{eqn:DI}) 
is quadratic and the minimization condition ${\partial
I}/{\partial \Psi_{\bf n}}=0$ yields a system of \emph {linear}
equations
$A{\bf \Phi} ={\bf b}$
where 
the vector ${\bf b}$ describes imposed boundary values and charge
distributions.  
Because $A$ is a sparse, symmetric and block-tridiagonal matrix,
the system can be solved efficiently by an iterative procedure,
which may, however, suffer from conditioning problems due to the
nonuniformity of the grid.
For an accurate solution, a mesh with high enough resolution is
required in regions where $\Phi({\bf r};s,V)$ varies strongly. We
used the PARDISO package~\cite{Schenk07} to solve the resulting
huge system of equations. An implementation of our FDM is
distributed under GNU-GPL license as the CapSol code~\cite{code}.

Once $\Phi({\bf r},s,V$=$1)$ is determined for several separations
$s$, the system capacitance is obtained as $C(s)= \int
\epsilon({\bf r}) |{\bf \nabla} \Phi |^2 \, d{\bf r} \simeq
\sum_{\bf n} \epsilon_{\bf n} \left| \nabla \Phi\right|_{\bf n}^2
\,  v_{\bf n}$. Then a simple second order FD approximation is
used to evaluate $C'(s)$ and $C''(s)$ from $C(s)$. The
electrostatic force acting on an arbitrary area $S$ of a
conducting part can also be evaluated as
\begin{eqnarray}\label{eqn:sigma}
{\bf F}_S = \frac{1}{2\epsilon_0}\int_S  \sigma(s)^2 {\hat n} dS,
\end{eqnarray}
where $\sigma(s) =-\epsilon {\partial \Phi}/{\partial n}$ is the
 surface charge density guaranteeing that the tip surface is an
equipotential, and
 $\hat n$ is the unit vector normal to the surface element $dS$.
 For a
system with cylindrical symmetry the net force on 
a part of the probe delimited by two cylinders of radii $\rho_1<\rho_2$ is
vertical and given by $F = \pi\epsilon_0 \int_{\rho_1}^{\rho_2} |\nabla\Phi |^2
\rho d \rho$, however we prefer to use Eq.~(\ref{eqn:FM})
 to calculate the total macrosocopic force on the probe.
In the following subsections we validate the performance of our
FDM by comparisons with previous results obtained by other methods
for cylindrically symmetric systems.
  We mainly consider the macroscopic model system described
in the caption of Fig.~\ref{fig:map} which shows 2D ($\rho,z$)
maps of the electrostatic potential computed by our FDM at three
magnifications differing by five orders.
  The conducting probe consists of a conical tip terminated by a spherical cap of radius
$R$ attached to a cantilever modelled as a disk  of the same area
as a typical cantilever,\cite{Jacobs98} and the sample by a thick
dielectric slab. Dirichlet boundary conditions are applied on a
very large cylindrical box. Note that the grid spacing changes by
six orders of magnitude
(hundredths of nm around the tip apex to tens of $\mu$m near the box walls).
 The indented contours in the bottom
right inset reveal the resolution of the finest grid, i.e. 0.02
nm. The contours in both magnified insets clearly show that for a
separation of 1 nm a large fraction of the voltage drop occurs
within the dielectric sample. Whereas the contour spacing between
the tip apex and the surface is constant to a good approximation,
 it gradually increases inside the dielectric, in contrast to what occurs in a parallel plate capacitor. Actually the
capacitance remains finite for an infinitely thick sample even
in the (macroscopic) contact limit $s \to 0$ (see Appendix
\ref{app:sphere}).

\subsection{\label{sec:test}{Convergence and Accuracy}}
\begin{figure}[h]
\begin{center}
\includegraphics[angle=-90,width=.35\textwidth]{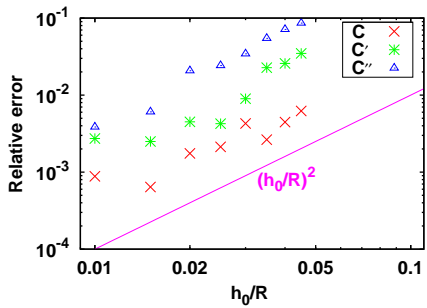}
\caption{\label{fig:err_h} (color online) Convergence analysis
with respect to the finest grid spacing $h_0$ for a conducting
sphere of radius $R$ in front of a 1 mm thick dielectric of
relative permittivity $\epsilon/\epsilon_0=5.9$. Points computed
by our FDM for the macroscopic capacitance $C$, the force $\propto
C'$ and force gradient $\propto C''$ are compared to the analytic
solution for a semi-infinite dielectric described in Appendix
\ref{app:sphere}. The sphere-surface separation is $R/20$ and the
computational box extends to $10^6R$ in the radial and vertical
directions. The straight line in the log-log plot indicates the
expected quadratic error scaling.} 
\end{center}
\end{figure}

\emph{Grid spacing:}
We first test our implementation for the problem of a conducting
sphere of radius $R$ 
separated by $s$ from a
semi-infinite dielectric surface for which an analytic solution of
controllable accuracy is available (see
Appendix~\ref{app:sphere}).
 A convergence analysis yields the parameters needed to achieve a
desired accuracy.
 Compared to the analytic solution, the
convergence with respect to the finest grid spacing $h_0$ of the
sphere-dielectric system (Fig.~\ref{fig:err_h}) shows a nearly
quadratic error scaling for small separations as expected for a
second order FDM. The accuracy could be improved by using a higher
order approximation for the electric field over further
neighboring points. However, a corresponding improvement of the
approximation of curved surfaces on the orthogonal FD-mesh is then
also required. 
 Note that, for consistency, the surface of the sphere must be
approximated as a staircase with variable step heights and widths
which also change when the grid-spacing is changed.  Thus the
error scaling deviates somewhat from the ideal straight line and
 is no longer quadratic when the separation increases. The
capacitance, force and force-gradient of this test system at a
rather small separation of $s=R/20$ can be calculated within a
relative error of 0.005 compared to the analytic solution if
$h_0=R/100$. For larger separations, this accuracy is
achieved even with a larger $h_0$. 

For the cantilever-tip-sample system shown in Fig.~\ref{fig:map}
and the results in the next subsection, a uniform grid with
$h^{(\rho)}=h^{(z)}=h_0=R/100$ is used around the tip apex up to a
distance of twice the tip apex radius in both radial and vertical
directions. Outside this range the grid becomes gradually coarser
with a growth factor of 1.01. In order to consistently preserve
the shape of the tip approximated by the orthogonal mesh, the
tip-sample separation must be changed in steps of $h_0$.

\begin{figure}
 \begin{center}
 \includegraphics[angle=-90,width=.45\textwidth]{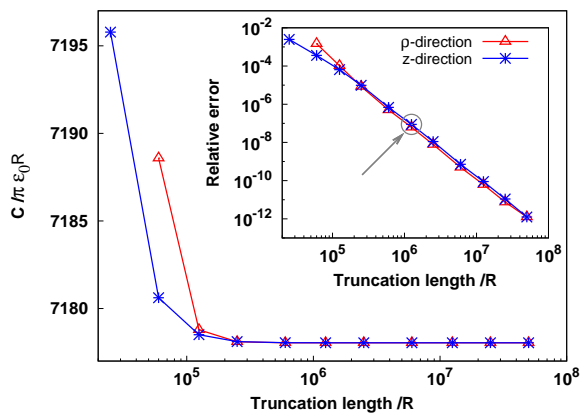}
 \caption{ \label{fig:err_L_max} (color online)
Convergence analysis with respect to the radial and vertical
extents of the FDM computational box for the macroscopic system
described in the caption of Fig.~\ref{fig:map}, the tip-sample
separation and finest mesh size being $s=R/20$ and $h_0=R/100$,
respectively. The normalized capacitance of the system approaches
the same asymptotic value upon increasing the truncation length in
one direction while the other one is sufficiently large and fixed.
Relative deviations with respect to the asymptotic value are shown
in the inset. The arrow indicates the truncation length adopted
in subsequent FDM computations.
 }
\end{center}
\end{figure}

\emph{Space truncation:} A convergence analysis with respect to
the size of the computational cylinder is shown in
Fig.~\ref{fig:err_L_max} for the model system described in
Fig.~\ref{fig:map}. The capacitance approaches the same asymptotic
value when the truncation length in a particular direction is
increased while the other one is kept fixed and sufficiently
large. If the computational box extends to $10^6R$ in the radial
and vertical directions, the relative deviation of the capacitance
from its asymptotic value is only $10^{-7}$ (as indicated by the arrow
in Fig.~\ref{fig:err_L_max}). We use these cutoff parameters
in all subsequent FDM computations reported here.

\begin{figure}
 \begin{center}
 \includegraphics[angle=-90, width=.5\textwidth]{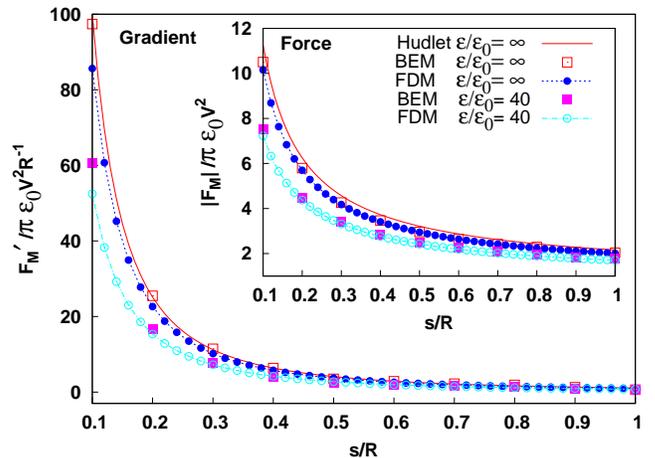}
 \caption{\label{fig:vsBEM} (color online)
Normalized macroscopic electrostatic force (inset) and
force-gradient computed by our FDM versus the normalized tip
separation $s/R$ from a dielectric ($\epsilon/ \epsilon_0=40.0 $)
and a conducting ($\epsilon/ \epsilon_0=\infty$) sample compared
to BEM computations (Ref.\cite{Shen08a}), as well as to Hudlet's
approximation (Ref.~\cite{Hudlet98}) in the second case (see
text). The cantilever is absent, as assumed in those two
treatments, but the remaining parameters are as described in the
caption of Fig.~\ref{fig:map}.}
 \end{center}
 \end{figure}

\emph{Comparison:}  In Fig.~\ref{fig:vsBEM} we compare results
obtained by our FDM with previous accurate BEM computations
~\cite{Shen08a} for a system like in Fig.~\ref{fig:map} but
\emph{without the cantilever}
for a conducting and a dielectric ($\epsilon/\epsilon_0=40$)
sample.
 The force and the force-gradient evaluated by
the two methods are in very good agreement for both kinds of samples.
For the conducting sample, Hudlet's analytic
approximation~\cite{Hudlet98}
deviates by only a few percent from the numerical results.  In the
following subsection we show that the contribution of the
cantilever can be quite appreciable for a dielectric sample.

\subsection{\label{sec:results}{Results}}

\begin{figure}
\begin{center}
\includegraphics[angle=-90, width=.5\textwidth]{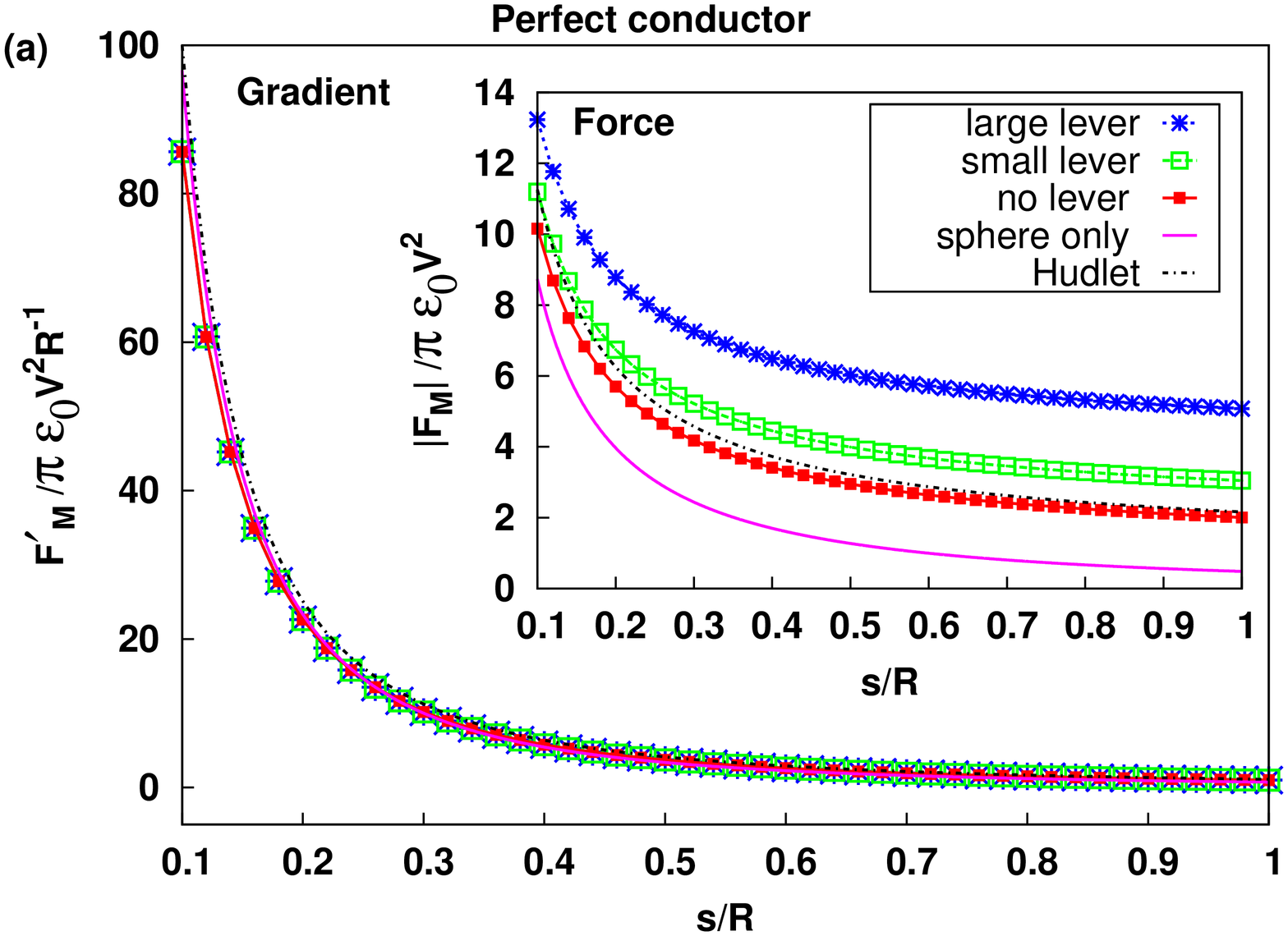}
\includegraphics[angle=-90, width=.5\textwidth]{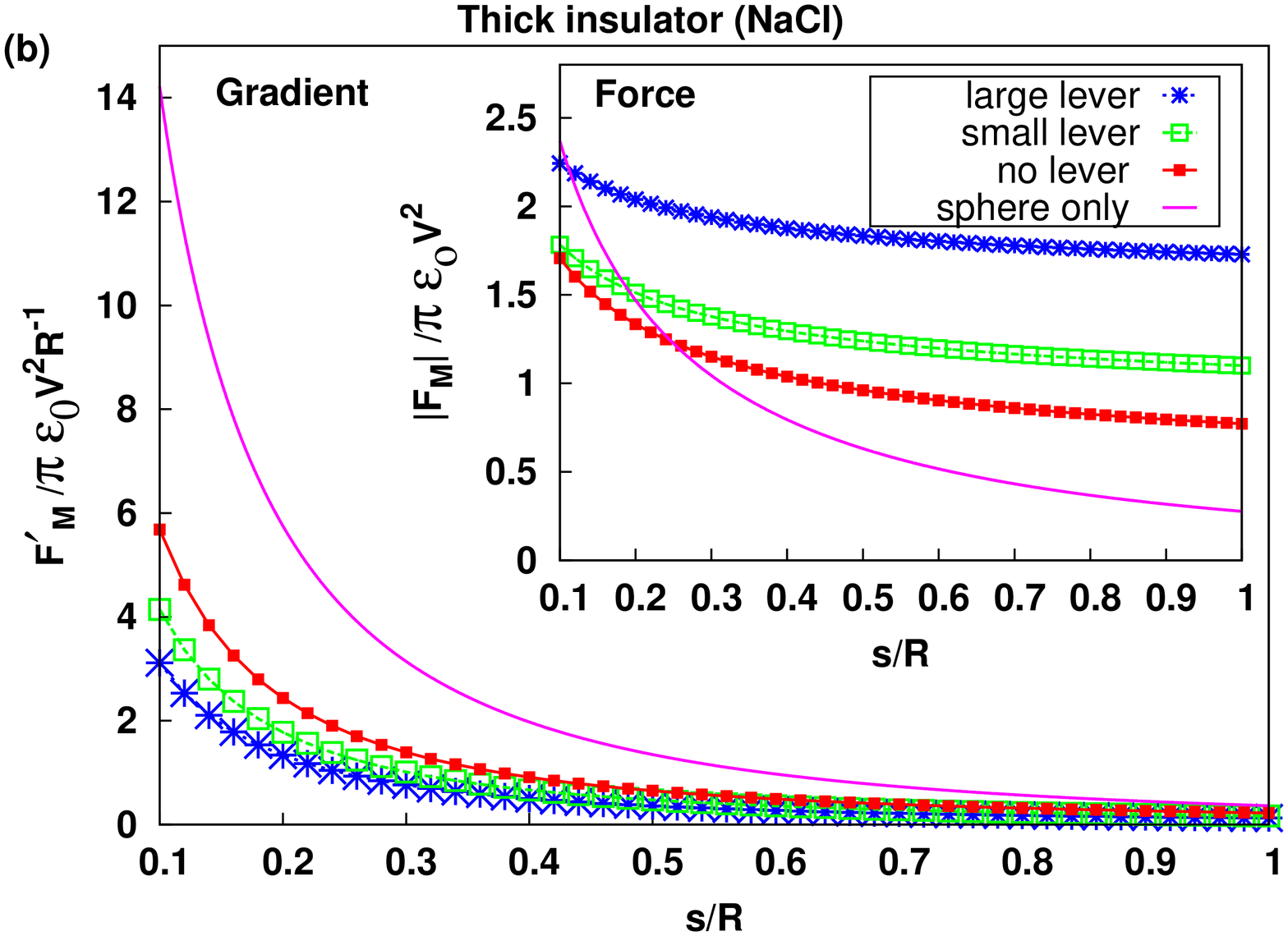}
\caption{\label{fig:f} (color online) Effect of the cantilever (size) on
the macroscopic electrostatic force (inset) and force-gradient at
different normalized tip separations from a conducting (a) and
dielectric (b) sample. The cantilever is modelled as either a
small or a large disk
with radii of 20 and 35 $\mu$m, respectively. Other parameters are
as in caption of Fig.~\ref{fig:map}. The solid lines show
corresponding results for a tip approximated by a conducting
sphere with radius $R=20$ nm obtained by summing the analytic
series for semi-infinite samples of both kinds (see Appendix
\ref{app:sphere}).}
\end{center}
\end{figure}

The macroscopic electrostatic force and force-gradient versus the
normalized tip-surface separation $s/R$ for the system in
Fig.~\ref{fig:map} are shown in Fig.~\ref{fig:f} for three
different geometries: without, with a small and a large cantilever
modelled as disks of thickness 0.5 $\mu$m. The small disk radius
is equal to the width of a typical rectangular AFM cantilever
(20$\mu$m) while the total area of the large disk (of radius 35 $\mu$m)
matches the area of the rectangular cantilever. The presence of
the cantilever increases the capacitance and the electrostatic
force. Because the cantilever is more than 10 $\mu$m away from the
surface, its contribution to the force is often considered
constant for tip-sample separations smaller than $R$, and
therefore does not contribute to the force gradient. Our
calculations [Fig.~\ref{fig:f}(a)] confirm that this is in fact
true for a conductive sample. In this case, the main contribution
to the force-gradient comes from the spherical cap, as can be seen
from the solid line which corresponds to the analytic solution for
a conducting spherical tip (see Appendix \ref{app:sphere}).
However, the conical shank of the tip and the cantilever affect
the force at large separations, as shown in the inset and noticed
earlier.~\cite{Hochwitz96,Belaidi97,Hudlet98,Jacobs98} On the
other hand, if $s/R$ is small, as shown in Fig.~\ref{fig:f}(b) and
also emphasized in previous work~\cite{Gomez01b,Sacha04,Sacha07},
\emph{over a thick dielectric sample both the force and the
force-gradient are significantly decreased, owing to field
penetration into the sample}.

\begin{figure}
\begin{center}
\includegraphics[angle=-0,width=.5\textwidth]{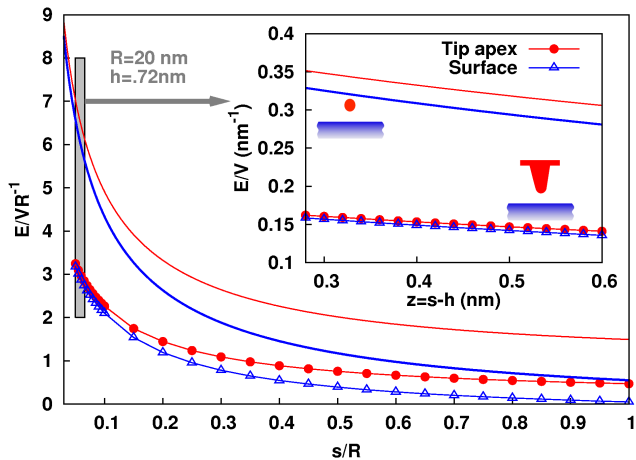}
\caption{\label{fig:Eofs} (color online) Normalized macroscopic
electric field just below the tip-apex and just above the
dielectric sample surface ($\epsilon/\epsilon_0$=5.9) versus their
normalized separation for the probe described in the caption of
Fig.~\ref{fig:map} (curves with symbols) and for a tip approximated by a conducting
sphere of the same radius ($R=20$~nm) (continuous curves).
Inset: zoom into the range where atomic-scale contrast appears;
the electric field between the tip and the surface changes by only
a few percent and is hence nearly uniform. }
\end{center}
\end{figure}

\begin{figure}
\begin{center}
\includegraphics[width=.4\textwidth]{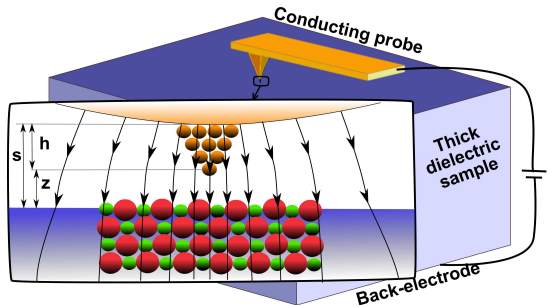}
\caption{\label{fig:afm} (color online) Sketch of the AFM setup
showing its macroscopic and microscopic parts on relevant scales.
The macroscopic tip-sample separation is $s=z+h$, where $h$ is the
nanotip height and $z$ is the nominal distance (without
relaxation) between the apex atom and the surface. The macroscopic
electric field $\bf {E}$ obtained by solving Poisson's equation is
applied as an external field to the atomistic subsystem shown in
the zoom window. In the range $z<$ 0.6 nm where atomic-scale
contrast appears, the electric field in the vacuum gap can be
considered to be uniform and equal to $E_z$ at the mid-point $s/2$
on the symmetry axis.}
\end{center}
\end{figure}

A quantity of particular relevance in our multi-scale approach is
the macroscopic electric field in the vacuum gap between the
spherical tip end and the sample surface
 which polarizes the microscopic system.  The variation of the
electric field normalized to $V/R$ at two points on the symmetry
axis just below the tip and just above the surface is shown in
Fig.~\ref{fig:Eofs} versus their normalized separation. The same
quantities shown magnified in the inset for nanotip separations
relevant for atomic-scale contrast, i.e. $z=s-h \lesssim$ 0.6~nm,
differ little and drop only weakly with increasing $z$.
In the same distance range
 the $z$-component of the electric field is two orders of magnitude
stronger than the radial component parallel to the surface. These
features are also clearly illustrated by the essentially
equispaced horizontal equipotential contour lines in the vacuum
region shown in the bottom inset of Fig.~\ref{fig:map}. This
important observation greatly simplifies the desired coupling to
atomistic calculations: we \emph{can consider the electric field
$E_z$ at the midpoint of the macroscopic tip-surface distance
$s=z+h$ as a uniform external field acting on the isolated
microscopic tip-sample system.} The connection between those two
scales is schematically illustrated in Fig.~\ref{fig:afm}.


Figure~\ref{fig:f} shows that
 for a conducting sample the force gradient can be accurately
described by a spherical tip if $s < R$, although the force
itself is increasingly underestimated at larger separations~\cite{Hudlet98,Guggisberg00}.
  In contrast, \emph{for a thick
dielectric sample}, the same description only provides the order
of magnitude of $F_M$ at small $s/R$, but exhibits a faster
decrease with increasing separation and overestimates $F'_M$.
Figure~\ref{fig:Eofs} reveals that a spherical model tip
overestimates the electric field $E_z$ under the tip
 at all separations, which then approaches $V/R$ 
on the sphere (and zero on the surafce) when $s \gg R$.
This occurs because the induced surface charges can spread to the
conical shank and the cantilever in the more realistic model. The
contributions of those parts to the force $F_M$ become
nevertheless stronger than that of the sphere alone already at
small $s/R$.
 In general, if $s/R \to 0$, the \emph{electric field under the
tip, hence the force and the force gradient are enhanced} owing to
an increasingly \emph{localized surface polarization} of both tip
and sample, but remain finite if the sample is a dielectric, as
explicitly demonstrated by the solution for a spherical tip (see
Appendix \ref{app:sphere}). Comparison with that solution (the
solid curves in Fig.~\ref{fig:f}) shows that \emph{even at small
separations} contributions from \emph{both the conical shank and
the cantilever contribute to the force}, whereas \emph{mainly the
conical shank affects the force gradient}. Hence, ignoring those
contributions causes an overestimation of the force-gradient if
the sample is an insulator.

\section{\label{sec:micro}  short-range electrostatic forces}
\subsection{\label{sec:previous} Evidence and previous models}
When an AFM tip approaches a surface, short-range forces
contribute to the tip-sample interaction and give rise to
atomic-scale contrast in NCAFM. Hereafter, ${F}_{\mu}$ denotes the
short-range force component perpendicular to the surface which can
be extracted from measurements of $\Delta f_1$ vs. the closest tip
approach distance $d$ in an oscillation cycle.\cite{Duerig99, Sader04}
The contrast observed in $V_{LCPD}$ in the same distance range cannot only be due to
the long-range electrostatic force,
but must be due to a short-range bias-dependent force. Arai and Tomitori
were the first to infer the existence of such a force from $\Delta f_1(V_b)$ curves recorded with
a cleaned and sharpened silicon tip closer than $0.5$ nm to a 7$\times$7 reconstructed Si(111)
surface.~\cite{Arai04}
 In particular, above a Si adatom, they found a narrow peak growing
with decreasing $d$ superposed on the usual parabolic dependence
around the plotted minimum of $-\Delta f_1(V_b)$ in their Fig.~1, i.e. for $V_b \simeq
V_{CPD}$.  Later the same authors pointed out that an even sharper
peak appeared at the same bias in the simultaneously recorded
tunneling current.~\cite{Arai06} This seemingly supported their
original suggestion that the additional attractive force causing
the peak in $-\Delta f_1(V_b)$ arose from the increased overlap
due to the bias-induced energetic alignment of dangling bonds
states localized at the tip apex and on Si surface adatoms. The
formation of a covalent bond between those states has been shown
to be responsible for the observed NCAFM contrast on the
7$\times$7 reconstructed Si(111) surface.~\cite{Lantz01} In
extensive recent measurements on the same system, however,
Sadewasser et al reported parabolic $\Delta f_1(V_b)$ curves, but
detected a rapid drop by about - 1V followed by a gradual increase
in $V_{LCPD}$ above a Si adatom with decreasing $d$ in the range
where the extracted short-range force showed a similar
behavior.~\cite{Sadewasser09} 
The apparent discrepancy with
respect to Arai and Tomitori's observations is not so surprising
because tunneling is seldom observed with clean silicon tips, 
although it is routinely measured in STM, as well as in NCAFM on
conducting and even semiconducting samples when using 
metal-coated silicon tips.~\cite{Sugimoto10,Kinoshita11} 
An appreciable position- and
distance-dependent tunneling current is, however, undesirable in
dynamic atomic-scale LCPD measurements because it violates the
basic assumption of an in-phase response to the AC voltage
modulation by causing phase shifts which are difficult to
compensate. This problem does not arise with insulating samples.
Nevertheless, Arai and Tomitori's basic idea that bias-induced
alignment of spatially localized surface states can lead to an
enhanced site-dependent attractive force remains plausible even if
a DC tunneling current cannot be sustained.
Thus Krok and coworkers~\cite{Krok08} suggested that the lower
$LCPD$ which they found across protruding In rows on the
c(2$\times$8) reconstructed InSb(001) surface was due to a
bias-induced local electron transfer from a polar dangling bond on
the electronegative Sb atom presumably picked by the Si tip to the
nearest electropositive surface In atoms. The same authors also
showed that the $LCPD$ contrast between different lateral
positions
decays exponentially with
increasing $d < $ 1nm.

Whereas bias-induced electron transfer is plausible for
narrow-bandgap semiconductors like those previously mentioned, it
is unlikely for overall neutral cleaved (001) surfaces of
wide-bandgap insulators like alkali halides which neither have gap
states, nor are reconstructed, but are only weakly
rumpled.~\cite{deWette85}  In Ref.~\cite{Bocquet08} the
atomic-scale $LCPD$ contrast observed on KBr(001) was attributed
to opposite surface cation and anion displacements in response to
local electric fields induced by the macroscopic (in accordance with our definition) field.
However, the authors approximated $E_z$ by the electric field
$V/R$ at the surface of an isolated conducting spherical tip,
the local unit cell polarizability by the bulk crystal
(Clausius-Mosotti) expression, and neglected the macroscopic surface
polarization.  Although essentially constant on the scale of a
nanometer-size nanotip, the latter, together with $E_z$ is
actually nonuniform on a lateral scale of order $\sqrt {Rs}$ for
separations $s \ll R$.
 They 
evaluated the macroscopic and microscopic surface charges
densities $\sigma_m$ and $\sigma_{\mu}$ induced on a conducting
model tip by their $E_z$ and by the displaced surface ions,
respectively.
 Using Eq.~(\ref{eqn:sigma}) they computed the modulation
of the electrostatic force.
 After further justified approximations, they obtained opposite LCPDs above cations and
anions which \emph{increased exponentially} with $d$. In a
subsequent article,~\cite{Nony09a} the same authors added a
macroscopic force roughly representing the interaction of the
cantilever with the back electrode, but still obtained a
surprisingly large maximum in the absolute LCPD for $d \simeq$~0.6~nm.
 In a subsequent publication~\cite{Nony09b},
even more reliable results were obtained  for a 
cubic NaCl cluster partly embedded into a conducting sphere
interacted with a NaCl(001) sample similar to ours via empirical
shell-model potentials. Cluster ions inside the sphere were fixed
and allowed to interact with the protruding cluster ions which
thus formed a nanotip with a net charge $+e$ at the apex. The
justification for such a model is that Si tips often pick up
sample material and that simulations based on the same code
produced reasonable results when compared to NCAFM measurements on
alkali halides, e.g. on KBr(001).~\cite{Hoffmann04,Ruschmeier08}
The results obtained can be considered representative of what is
expected for a
 strongly polar tip interacting with an ionic crystal. Recently a
simpler model for such a tip (conducting sphere terminated by a
point charge or a point dipole)
 could account for the observed variation of the LCPD over
a few nanometers.~\cite{Barth10}

Earlier studies mentioned that the short-range tip-sample
interaction is bias-dependent but provided no recipe to
investigate it theoretically.  Moreover, they did not clarify how
long-range and short-range bias-dependent forces are connected and
the role of each in the observed KPFM signals.
 In the following sections we answer all of these questions and
obtain and analyze in detail theoretical expressions for the
site-dependent LCPD. Our approach is not limited to particular
materials, but results are presented for the system described in
the following Section which is representative of a neutral, but
polarizable reactive clean Si tip interacting with an ionic crystal.

\subsection{\label{sec:DFT} Density functional computations}
As illustrated in Fig.~\ref{fig:afm} our microscopic system
consists of a nanotip of height $h$ protruding from the spherical
end of the macroscopic tip
 and of a wider two-layer slab of sample atoms.  Computations are
performed within the local-density approximation to density
functional theory (DFT) using norm-conserving HGH pseudopotentials~\cite{HGH}
and the BigDFT package.~\cite{BigDFT}  Relying on a
wavelet basis set with locally adjustable resolution, this package
calculates the self-consistent electron density, the total energy
and its electrostatic component with selectable boundary
conditions~\cite{PoissonFBC} i.e. periodic in two directions and free in the third
in our case. This allows us apply an external field perpendicular
to the surface without artifacts which can arise from periodic images
in the $z$ direction when using plane-wave of mixed basis sets.
As already explained, the voltage biased macroscopic system
determines the uniform electric field $E_z \propto V=V_b-V_{CPD}$
applied into the microscopic part (see Fig.~\ref{fig:afm}). This
provides the desired well-defined relationship between the
bias-voltage and short-range forces which was lacking in previous
approaches to LCPD contrast based on DFT
computations.~\cite{Sadewasser09,Masago10}

\begin{figure}
\begin{center}
\includegraphics[width=0.35\textwidth]{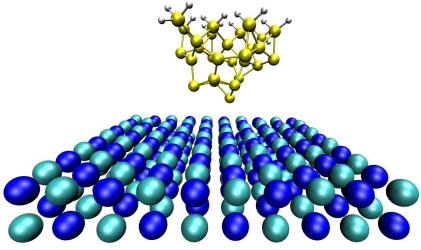}
\caption{\label{fig:posinp} (color online) The microscopic Si-tip
NaCl-slab system used in our \emph{ab initio} DFT calculations.
The apex of a silicon AFM tip is modelled as a pre-relaxed
$Si_{29}H_{18}$ cluster.
All eight atoms in the top (001) layer are passivated by hydrogen
atoms and kept fixed.
The position of the foremost $Si$ atom is $(x,y,z)$, $z$ being the
height from the surface. The model sample consists of two
 NaCl(001) layers each containing 10$\times$10 ions with
the bottom layer kept frozen. 
Periodic boundary conditions are applied along the $x$ and $y$
directions.}
\end{center}
\end{figure}

Figure~\ref{fig:posinp} illustrates the microscopic system used in
the DFT computations reported here.  The nanotip at the very end
of a silicon tip is modelled as a cluster with a fixed (001) base
of eight Si atoms with all dangling bonds
 passivated by H atoms in order to mimic the connection to the rest of the tip.
The remaining Si atoms were pre-relaxed using the Minima Hopping
Method~\cite{Goedeker04} previously employed to generate low-energy structures of
silicon clusters and of similar model tips.~\cite{Ghasemi08} As in
that work, the free Si atoms adopted a disordered configuration
with several exposed under-coordinated atoms. In particular the
protruding apex atom is threefold coordinated and hence has a
dangling bond with a small dipole moment pointing towards the
surface. 
As we verified, 
a distance five times the lattice constant of NaCl is large
 enough to get rid of  
 the electrostatic interaction between this nano-tip and its images
in the periodic directions. 
Therefore our sample consists of two 10$\times$10 NaCl(001) layers containing 
200 ions in total.
For such a large system, we evidently perform calculations
only at one single k-point, namely 
center of the surface Brillouin zone.
With periodic boundary conditions applied along the main
in-plane symmetry directions pre-relaxation of the sample only
caused a small rumpling which preserved the basic periodicity of
the truncated (001) surface.
 Although the silicon model tip and
the sample were initially individually pre-relaxed, all tip and
sample atoms were subsequently frozen in some of our KPFM
simulations. In this way we could assess pure electronic
polarization effects without effects due to the
interaction-induced displacements of nuclei.

\begin{figure}
\begin{center}
\includegraphics[angle=-90,width=.45\textwidth ]{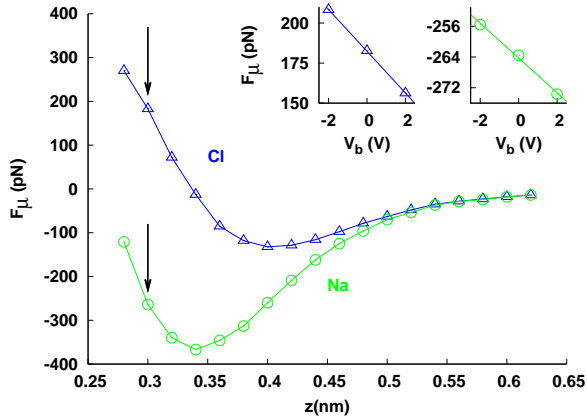}
\caption{\label{fig:Fmu} (color online) Microscopic force on the
Si nanotip above Na and Cl surface ions from \emph{ab initio}
calculations without an applied electric field. Insets: variation
of the force as a function of the macroscopic bias voltage at a
tip-surface separation of 0.30 nm. The error-bar (not shown) is
$\sim1pN$. }
\end{center}
\end{figure}
{
The silicon model tip was positioned so that its foremost atom was
0.65 nm above a sodium and chlorine surface ion, then moved
towards the sample in steps of 0.02 nm. At each step the Kohn-Sham
equations are solved iteratively.
The topmost layer of the Si tip together with the
passivating H atoms, as well as the bottom layer of the slab are
kept fixed while other ions are free to relax
until the Hellman-Feynman force exerted on each ion is less than  1~pN.
This extremely tight tolerance is required only
when the relative variation of the force when the bias changes is very small.
 The force $F_{\mu}$ exerted on the model tip is obtained by summing the
$z$-components of the forces over atoms of the tip.
Since the free atoms are well relaxed, their contribution to
that force is not significant and was used as a measure of the error in forces.
 Figure~\ref{fig:Fmu} shows the
microscopic force versus the tip-apex separation from Cl and Na
surface sites without applied electric field. The same procedure
is repeated at each tip-sample separation for a few
field strengths $E_z$ determined as explained in
subsection~\ref{sec:results} for effective biases $-2 \le
V=V_b-V_{CPD}\le 2$ Volts applied to the macroscopic tip.
For such biases and distances where $F_{\mu}$ becomes
site-dependent, a nearly uniform macroscopic electric field 
 of $\sim$0.15~V/nm occurs in the vacuum gap,
 as discussed in Sec.~\ref{sec:macro} and illustrated in the inset of Fig.~\ref{fig:Eofs}.
No instabilites caused by electronic and/or atomic rearrangements
appeared in that range of parameters. The variation of the force
at the particular separation $z=$ 0.3~nm is shown in the insets in
Fig.~\ref{fig:Fmu}. In contrast to the macroscopic capacitive
force, \emph{the short-range force depends linearly on the applied
bias voltage}.
As explained elsewhere,~\cite{sadeghiMicroModel} this linear term
arises from the interaction between distance-dependent but
$V$-independent net charge densities on the tip and sample
surfaces with $V$-induced changes on the opposite surface
and with the macroscopic electric field.
Earlier studies obtained such a term by treating native ions or
charged atoms adsorbed on the sample surface and/or the tip apex
as point charges.~\cite{Bocquet08,Barth10,Bocquet11}
%
 Deviations from the linear behavior could occur for larger biases,
especially near instabilities, as observed in computations for a
charged nanotip.~\cite{Nony09b}

The \emph{basic quantity which determines the deviation of the
LCPD from the background CPD }is the voltage-independent slope of
the short-range force with respect to the applied voltage
\begin{eqnarray}
a(x,y,z)=\frac{\partial}{\partial V} F_{\mu}(x,y,z;{\bf E}(V))  \label {eqn:Fmu}.
\end{eqnarray}
As discussed in the Introduction, the background CPD is not a
well-defined quantity for an insulator.  For a real doped silicon
tip-NaCl(001) sample, it would be different from the CPD of our
microscopic system if charge equilibrium is achieved, as enforced
by the self-consistency of the computations.  Besides, no CPD is
explicitly included in the description of the macroscopic system.
Thus the effective bias $V=V_b-V_{CPD}$ would differ from that in
a real system. Nevertheless, as long as this bias is in the Volt
range, the slope $a$ is unaffected.

Fig.~\ref{fig:slope}(a) shows that the slope $a$ exhibits a
characteristic site-dependent distance dependence at separations
less than 0.5 nm, and is larger above the more polarizable Cl ion.
The underlying physics will be explained elsewhere.~\cite{sadeghiMicroModel} 
 The microscopic force-gradient $F'_{\mu}$ is also a linear function of bias
voltage, and
the distance-dependence of its slope $a'(x,y,z)=\partial
F'_{\mu}/\partial V$, approximated to second order by linear
interpolation between adjacent points on both sides of a given
$z$-value, is shown in Fig.~\ref{fig:slope}(c).
Figures~\ref{fig:slope}(b) and ~\ref{fig:slope}(d) show that $a$ and $a'$ are
weaker if relaxation is allowed but that contrast appears below
nearly the same distance and exhibits almost the same distance
dependence.  Thus, \emph{for the assumed neutral
Si nanotip, the contrast is mainly due to electronic polarization
rather than to bias-induced ion displacements}.
\begin{figure}
\begin{center}
\includegraphics[angle=-90,width=\columnwidth]{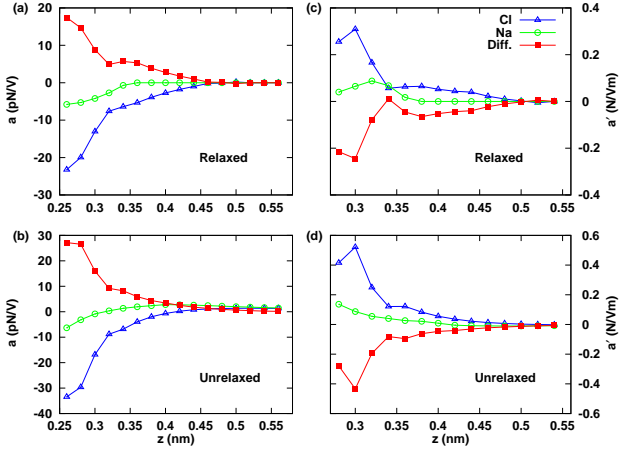}
\caption{\label{fig:slope} (color online) Distance-dependence of
the slopes $a=\partial F_{\mu}/\partial V$ and $a'=\partial
F'_{\mu}/\partial V$ above Na and Cl surface ions with (a,c) and
without(b,d) relaxation of the free atoms and ions during tip
approach. The difference (contrast) between Na and Cl sites is
shown by red (filled) symbols.}
\end{center}
\end{figure}

In the approximation that the  macro- and microscopic systems are
coupled only through the macroscopic electric field, the
$z$-component of the total force exerted on the tip is
\begin{eqnarray}\label{eqn:Ftot}
F=F_M(s;V) +F_{\mu}(x,y,z;V) +  F_{vdW}(s)
\end{eqnarray}
where $s=z+h$ and $V=V_b-V_{CPD}$. The long-range van der Waals
force $F_{vdW}$ is bias- and site-independent, being only a
function of the mesoscopic geometry and is therefore
henceforth ignored, although it affects the resonance frequency
shift $\Delta f$ in a NCAFM measurement.  The macroscopic force
$F_M$ is capacitive ($\propto V^2$) while the microscopic force
$F_{\mu}$ has been shown to be linear in $V$.

Three additional corrections couple the bias-dependent macro- and
microscopic forces. The first correction $\delta C'V^2/2$
 is due to an additional capacitive contribution caused by the presence of a
polarizable nanoscale object in the gap between the macroscopic
bodies. Owing to the small lateral dimensions of the nanotip
compared to the radius of the macroscopic tip end, this correction
is small,~\cite{Bocquet11,sadeghiMicroModel} although it can
become appreciable and site-dependent if the nanotip apex is
charged.~\cite{Nony09b}

The second correction arises only in that case or if the nanotip
has a large net dipole moment~\cite{Teobaldi11}.
This leads to a site-independent LCPD with an approximate
power-law approach towards a background CPD of several Volts.  The
interaction of
the nanotip charge distribution with the macroscopic field $\bf
E$ could in principle be included in our description at
separations $s$ where $\bf E$ can no longer be considered
uniform. In that range, however, the charge or dipole might be
approximated as point objects, as justified in the case of a
conducting sample in the Supplemetary Material of
Ref.~\cite{Barth10}. Because the charge or dipole are intrinsic,
the interaction is proportional to $V$, so that this correction would
give rise to long-range contributions to the slopes $a$ and
$a'$.~\cite{Barth10,Bocquet11} In the case of our neutral Si
nanotip, this correction is small.

The third correction
 arises because in reality the nanotip is in electrical contact
with the macroscopic tip, so that the electron density at the
interface differs from that near the top of our isolated silicon
cluster.
 However, this model tip is large enough,
 so that the charge distribution near the apex, which dominates
$F_{\mu}$ is not much affected, in contrast to models with smaller
model tips. The self-consistently determined microscopic electric
field between the nanotip apex and the sample surface differs from
the applied macroscopic field $E_z$, but this effect is already
included in the computed microscopic force.

\begin{figure}
\begin{center}
\includegraphics[width=.7\columnwidth ]{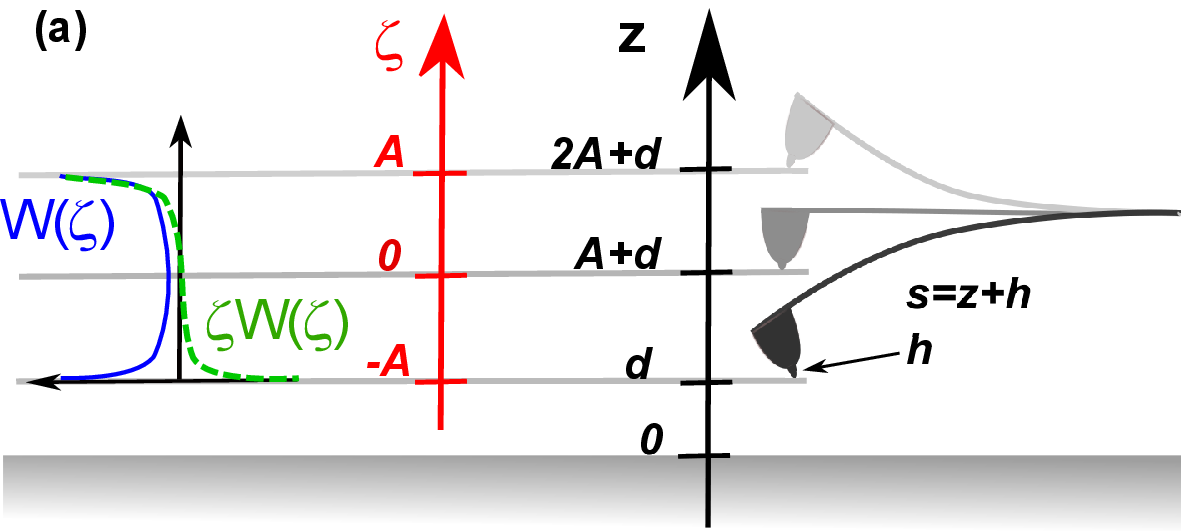}
\includegraphics[angle=-90,width=\columnwidth]{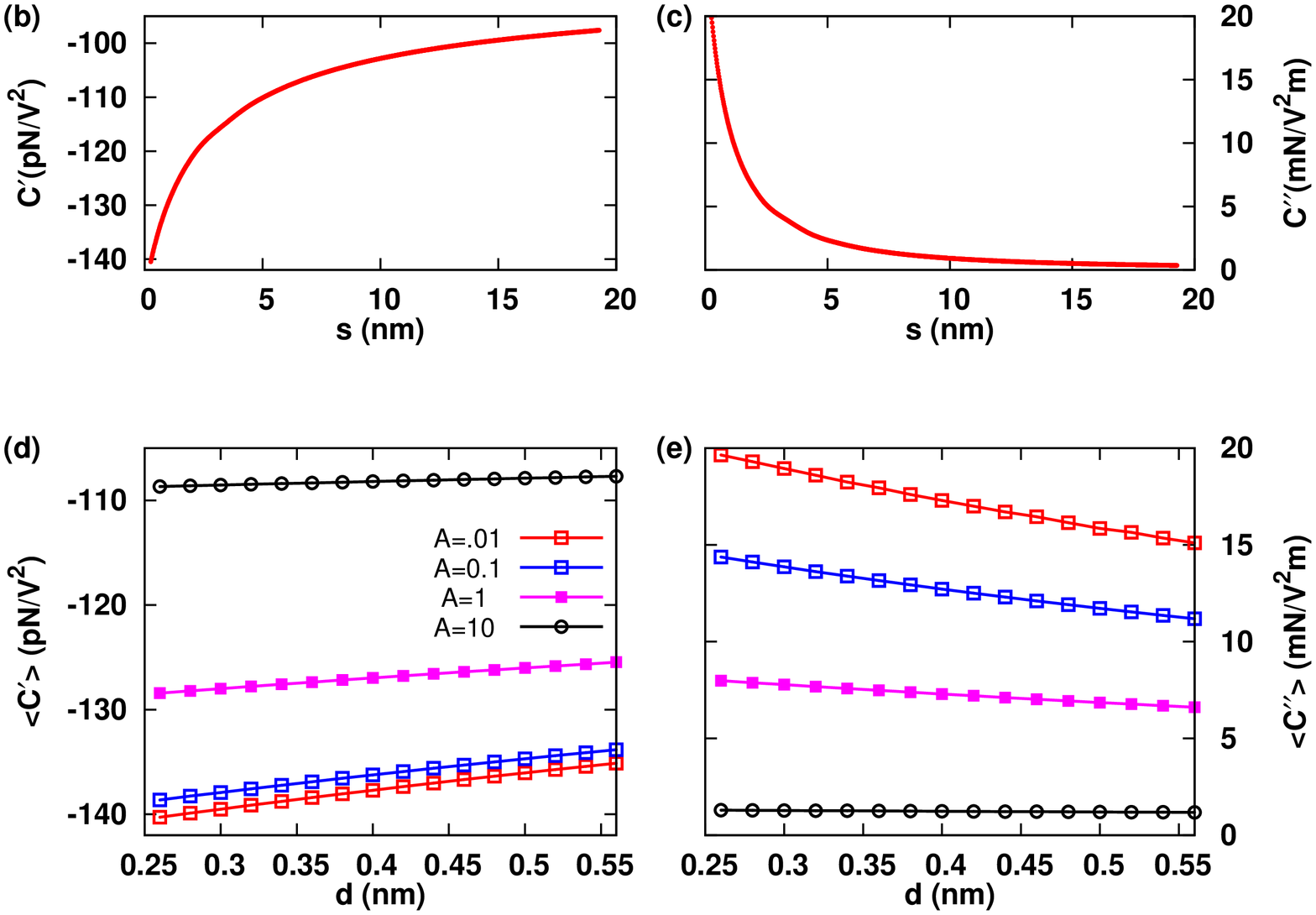}
\caption{\label{fig:CpCpp} 
(color online) (a) Sketch of the cantilever oscillating in its
fundamental mode of a tip with finite amplitude $A$ at the end .
The weight functions used for calculating the cycle averages in
Eqs.~(\ref{eqn:<g>}-\ref{eqn:<gp>}) are shown as functions of
$\zeta=z-d-A$ where $d$ is the closest tip apex-sample separation.
Dependencies of the first (b) and second (c) spatial derivatives
of the capacitance on the macroscopic separation $s = z+h$
calculated for the setup in Fig.~\ref{fig:map}, and of their cycle
averages (d,e) on $d$ for tip oscillation amplitudes $A=$0.01,
0.1, 1 and 10 nm.}
\end{center}
\end{figure}

\begin{figure}
\begin{center}
\includegraphics[width=.5\textwidth ]{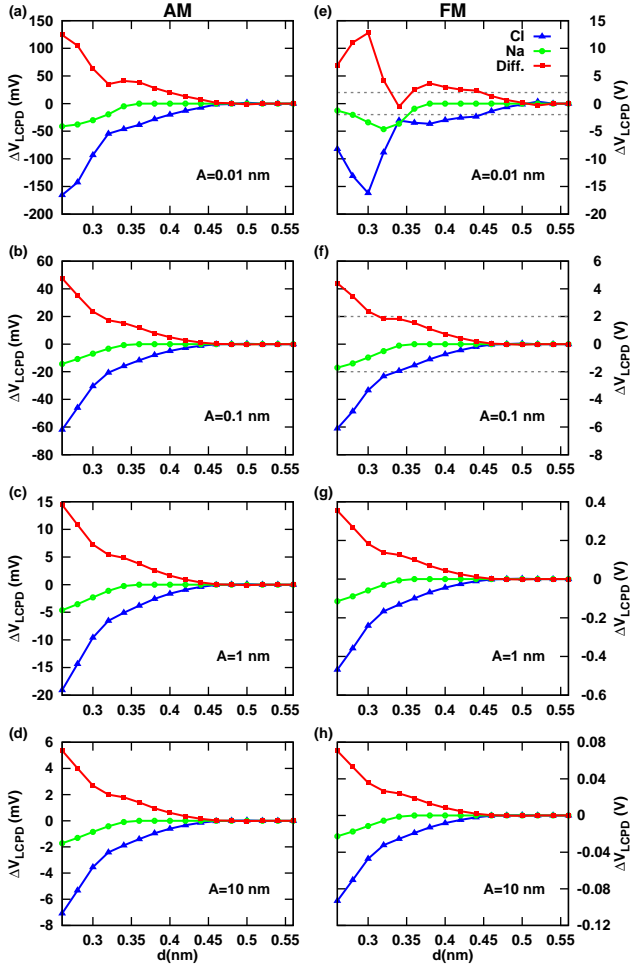}
\caption{\label{fig:LCPD} (color online) Calculated deviations
$\Delta V_{LCPD}$ for AM- (left column) and FM-KPFM (right column)
versus closest tip apex-sample for tip oscillation amplitudes
$A=0.01$ nm (a,e),$A=0.1$ nm (b,f), $A=1$ nm (c,g), and $A=1.0$ nm
(d,h).
In (e,f) the dashed horizontal lines indicate the range of validity of our DFT calculations ($\pm$2V).
 }
\end{center}
\end{figure}

\section{\label{sec:kpfm}AM and FM KPFM signals and Local Contact Potential Differences}
\subsection{\label{sec:zero A} Ultrasmall amplitude limit}

The force gradient is more sensitive than the force to short-range
interactions which are responsible for atomic-scale contrast in
NCAFM and KPFM.  Direct detection of the gradient is in principle
possible if the variation of $F_{\mu}$ over the peak-to-peak
oscillation amplitude is linear, e.g. if $2A$ is comparable to the
spacing 0.02 nm of the computed points in Fig.~\ref{fig:Fmu}. We
first consider this simple limit which is commonly assumed in the
KPFM literature, but is seldom achieved in NCAFM experiments.
 As explained in the Introduction, $V_{LCPD}$ is operationally
defined by nulling the KPFM signal generated by the force
component $F_{\omega}$ at the modulation frequency. Assuming that
the response $V_{AC}$ is linear and instantaneous, $F_{\omega} =
(dF/dV_b) V_{AC}$, and
the deflection signal detected in AM-KPFM would be proportional to
\begin{eqnarray}
F_{\omega} &=& \left[C'(z+h)\left(V_{DC}-V_{CPD}\right) +
a(x,y,z) \right]V_{AC} \label {eqn:Ff}
\end{eqnarray}
in the ultrasmall amplitude limit,
 and would be nulled if
\begin{eqnarray}
V_{DC}= V_{CPD} -\frac{a(x,y,z)}{ C'(z+h)}. \label{eqn:VLCPD}
\end{eqnarray}
 Because the background $V_{CPD}$ is not well-defined, and only
$a(x,y,z)$ is site-dependent, we consider only the deviation of
$V_{LCPD}$ from $V_{CPD}$ which is responsible for atomic-scale
contrast, i.e.
\begin{equation} \label{eqn:LCPDAM}
\Delta V_{LCPD}^{AM}=-\frac{a(x,y,z)}{C'(z+h)},
\end{equation}
 For a thick dielectric sample, as illustrated by Fig.~\ref{fig:slope}(a) and by the points for
$A$=0.01 nm in Fig.~\ref{fig:CpCpp}(d), the $z$-dependence of $C'$
is weak over the range where $a(x,y,z)$ is appreciable
($s=z+h<1$~nm). Therefore $\Delta V_{LCPD}$ differs from
$a(x,y,z)$ by an essentially $z$-independent factor.
 Depending on the nanotip height $h$, this may no longer hold in
the case of a
 conductive sample or thin dielectric film on a conductive
substrate.

In FM-KPFM the contribution of the modulated electrostatic force
component $F_{\omega}$ to the frequency shift of the first
resonant mode $\Delta f_1$ is detected and nulled. In the
ultrasmall amplitude limit $\Delta f_1$ is proportional to the
force-gradient\cite{Albrecht91} and would therefore be nulled if
\begin{eqnarray} \nonumber 
F'_{\omega} &=& \left[C''(z+h)\left(V_{DC}-V_{CPD}\right) +
a'(x,y,z) \right]V_{AC}=0 .
\end{eqnarray}
 The FM-counterpart of Eq.~(\ref{eqn:LCPDAM}) is therefore
\begin{eqnarray} \label{eqn:LCPDFM}
\Delta V_{LCPD}^{FM}=-\frac{a'(x,y,z)}{ C''(z+h)}.
\end{eqnarray}
Again, as illustrated by by Fig.~\ref{fig:slope}(c) and by the
points for $A$=0.01~nm in Fig.~\ref{fig:CpCpp}(e), the site- and
distance dependence of this deviation is determined
by $a'(x,y,z)$, but $\Delta V_{LCPD}^{FM}$ again differs from the
numerator by an almost $z$-independent factor. The calculated LCPD
deviations for $A = 0.01$~nm in the AM and FM modes are plotted in
Figs.~\ref{fig:LCPD}(a) and (e). Note that $\Delta V_{LCPD}^{FM}$
would be about hundred times stronger and would exceed the range
of validity  ($\pm$2~V) of our DFT computations 
(limited between the horizontal lines in Figs.~\ref{fig:LCPD}(e,f)), 
as well as the measured results in experiment,
hence cannot be trusted.
Therefore it is important to consider averaging over the
range covered by the finite tip oscillation.

\subsection{Finite amplitude expressions \label{sec:finiteA} }
In NCAFM with cantilevers the oscillation amplitude $A$ is
between several and a few tens of nanometers, so that the
macroscopic capacitive electrostatic force can change by several
orders of magnitude over an oscillation cycle. In practice,
 the detected AM and FM KPFM signals are given by differently
weighted averages, namely~\cite{Kawai10}
\begin{align*}
\left< F_\omega \right>_{} = \frac{1}{2\pi} \int_{0}^{2\pi} F_\omega[d+A(1+\cos\phi)]d\phi
\end{align*}
and~\cite{Giessibl97}
\begin{align*}
%
kA \frac{\left< \Delta f_\omega \right>_{}}{f}
 = - \frac{1}{2 \pi} \int_{0}^{2\pi} F_\omega [d+A(1+\cos\phi)]\cos \phi d\phi
\end{align*}
where $k$ is the flexural stiffness of the cantilever and
 $d=z_{min}$ is the closest tip apex-sample separation.
 Substituting the force from Eq.~(\ref{eqn:Ff}) and setting these
averages to zero, one obtains
\begin{eqnarray} \label{eqn:VL3}
\Delta V_{LCPD}^{AM}= -\frac{\langle a(x,y,z) \rangle}{\langle C'(z+h)\rangle},
\\ \label{eqn:VL4}
\Delta V_{LCPD}^{FM}= -\frac{\langle a'(x,y,z) \rangle}{\langle
C''(z+h)\rangle},
\end{eqnarray}
where the cycle averages defined as
\begin{eqnarray} \label{eqn:<g>}
\langle  g \rangle &\equiv& 
 \frac{1}{\pi}  \int_{-A}^{A} W(\zeta)g(d+A+\zeta) d\zeta,
\\
\label{eqn:<gp>}
\langle  g' \rangle&\equiv& 
 \frac{1}{\pi A}  \int_{-A}^{A} \zeta W(\zeta)g(d+A+\zeta) d\zeta
\nonumber \\ &=&
\frac{1}{\pi A^2}  \int_{-A}^{A} \sqrt{A^2-\zeta^2} g'(d+A+\zeta) d\zeta
\end{eqnarray}
depend both on $d$ and $A$. As depicted in
Fig.~\ref{fig:CpCpp}(a), $\zeta=z-(A+d)$
 whereas $W(\zeta)=(A^2-\zeta^2)^{-1/2}$ is a weight function with
square root singularities at the turning points of the
oscillation.  The expression on the second line of
Eq.~(\ref{eqn:<gp>}) justifies the notation $\langle  g' \rangle$
and shows that this quantity tends to $g'(d+A)$ when $A \to 0$,
besides helping to relate the distance dependence of $\Delta
V_{LCPD}^{FM}$ to those of $a'(x,y,z)$ and  $C''(z+h)$.
However, because $a(x,y,z)$ is computed with high precision,
whereas $a'(x,y,z)$ is obtained by interpolation, we use the
expression on the first line for numerical purposes. Furthermore,
since $a(x,y,z)$ is known only at equispaced separations $z_i$
where the DFT computations have been performed, the integrals in
Eqs.~(\ref{eqn:<g>}-\ref{eqn:<gp>}) must be discretized.  The adopted
procedure, which deals with the singularities of the weight
function $W(\zeta)$ at the integration limits,~\cite{Pfeiffer04}
is presented in Appendix~\ref{app:averaging}. There we also show
that the discretized version of the expression in the first line
of Eq.~(\ref{eqn:<gp>}) reduces to the second order FD approximation of
$g'(d+A)$ when $2A$ matches the spacing between adjacent $z_i$
values, in accordance with the expression on the second line.

\subsection{\label{sec:LCPD} Results}

Owing to the very different z-dependencies of $a(z)$ and
$C'(z+h)$, shown respectively in Figs.~\ref{fig:slope}(a) and
\ref{fig:CpCpp}(b), their cycle averages depend in different ways
on $d$ and $A$. The same holds for $a'(z)$ and $C''(z+h)$, shown
respectively in Figs.~\ref{fig:slope}(c) and \ref{fig:CpCpp}(c).
Figures~\ref{fig:CpCpp}(d) and ~\ref{fig:CpCpp}(e) show the cycle averages of $C'$ and
$C''$ versus the closest tip-apex approach distance $d$ for
oscillation amplitudes $A=$ 0.01, 0.1, 1 and 10~nm, whereas the
cycle-averages of $V_{LCPD}$
 calculated from Eqs.~(\ref{eqn:VL3},\ref{eqn:VL4}) are plotted in
Fig.~\ref{fig:LCPD} for AM-KPFM (left column) and FM-KPFM (right
column) for the same amplitudes in the range where $a(z)$ is
finite. In that range, the cycle averages for $A=$~0.01~nm agree
with the non-averaged quantities.
Since
 the primary quantities were calculated at points spaced by 0.02~nm,
 this is not surprising in view of the remarks at the end of
the preceding subsection.
 Thus, apart from small deviations introduced by the
discretization procedure, 
 the points in Figs.~\ref{fig:LCPD}(a) and \ref{fig:LCPD}(e)
which were actually calculated for $A=0.01$~nm
coincide with those
 given by Eqs.(\ref{eqn:LCPDAM},\ref{eqn:LCPDFM}), and exhibit
essentially the same distance dependencies as $a(d)$ and $a'(d)$,
as already discussed in the subsection~\ref{sec:zero A}.

Already above $A$=0.1~nm, however, the LCPD contrasts in both
modes exhibit almost the same spatial dependence as $a(d)$,
although their respective magnitudes decrease if $A$ is increased.
Nevertheless, $\Delta V_{LCPD}^{FM}$ significantly exceeds $\Delta
V_{LCPD}^{AM}$; this can be understood as follows. As seen in
Figs.~\ref{fig:CpCpp}(d) and \ref{fig:CpCpp}(e), $\langle C''\rangle$ drops much faster
than -$\langle C' \rangle$ if $A$ is increased.  As explained in
the discussion of Fig.~\ref{fig:f}(b) this behavior reflects the
increasing influence of the relative contributions of the tip
shank and of the cantilever to $C'(z+h)$ in the range covered by
the peak-to-peak oscillation. Especially $\langle C' \rangle$ is
affected by the cantilever contribution which causes the very
gradual levelling of $C'(z+h)$ apparent in
Fig.~\ref{fig:CpCpp}(b).  As seen in Fig.~\ref{fig:CpCpp}(c), this
slowly varying contribution tends to cancel out in $C''(z+h)$,
and, according to the second line in Eq.~(\ref{eqn:<gp>}), in
$\langle C'' \rangle$ as well.

On the other hand, $\langle a \rangle$ and $A \langle a' \rangle$ essentially coincide once $a$
exceeds the range where $a$ is noticeable. Indeed, the main
contributions to those averages come from the vicinity of $z=d$
where the integrands in Eqs.~(\ref{eqn:<g>}) and (\ref{eqn:<gp>})
(first line) match. Expanding $W(\zeta)$ about this turning point,
one finds that $\langle a \rangle \sim A^{-1/2}$ whereas $ \langle a' \rangle \sim A^{-3/2}$, just
like $\Delta f_1$ behaves in NCAFM.~\cite{Giessibl97} According
to Fig.~\ref{fig:CpCpp}(b,c) the same argument cannot be applied
to $\langle C'' \rangle$ for $A \le$ 10 nm, and not at all to $\langle C' \rangle$ because
$C'(s)$ varies only slowly up to $s = R = 20$ nm.
 Fig.~\ref{fig:ofA} shows how the finite oscillation
amplitude affects the
relevant cycle averages, as well as $\Delta V_{LCPD}$ in the AM mode
(left column) and in the FM mode (right column) at the closest
tip apex-sample separation $d=0.30$~nm indicated by arrows in
Fig.~\ref{fig:Fmu}.
\begin{figure}
\begin{center}
\includegraphics[angle=-90, width=\columnwidth]{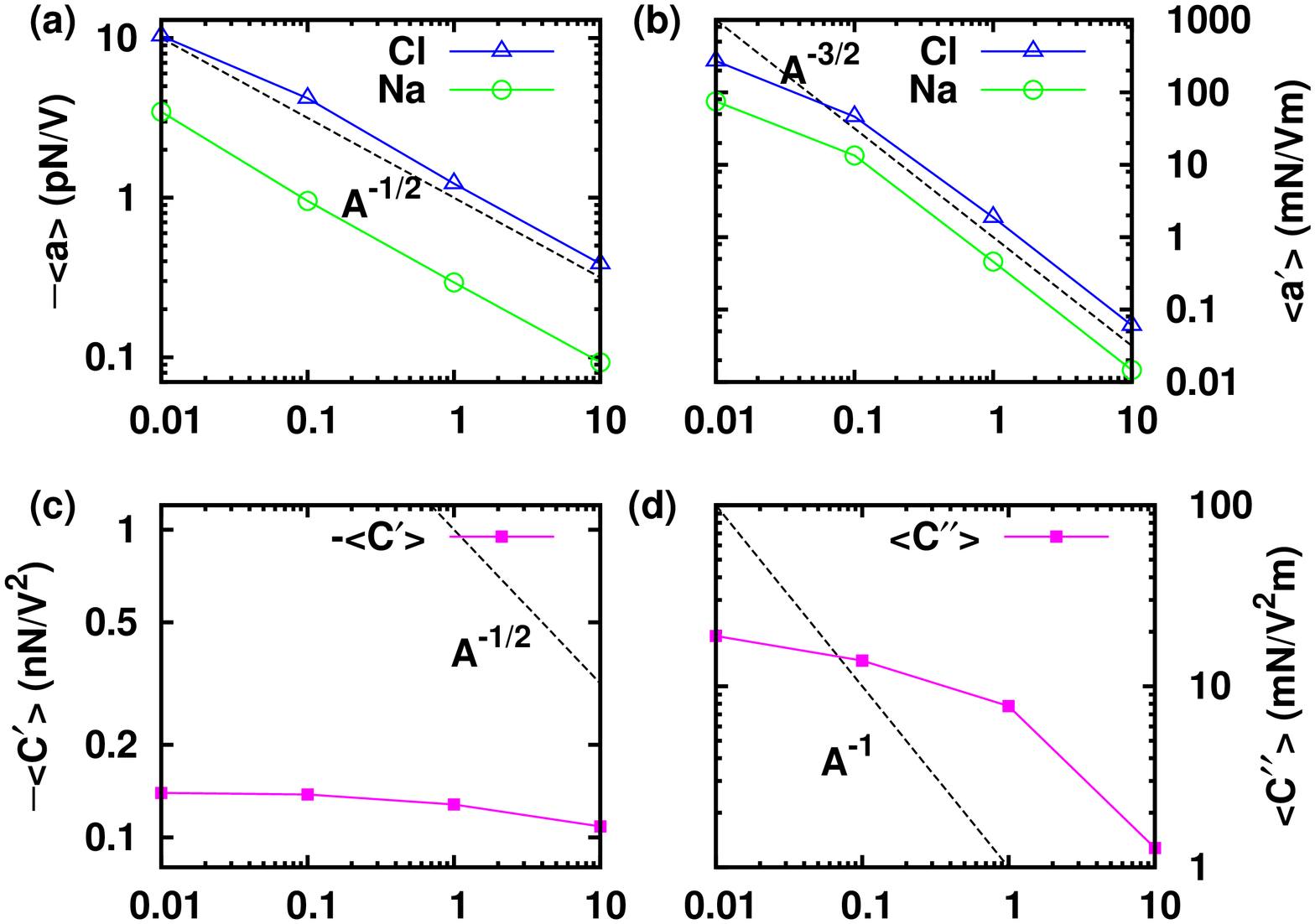}
\includegraphics[angle=-90, width=\columnwidth]{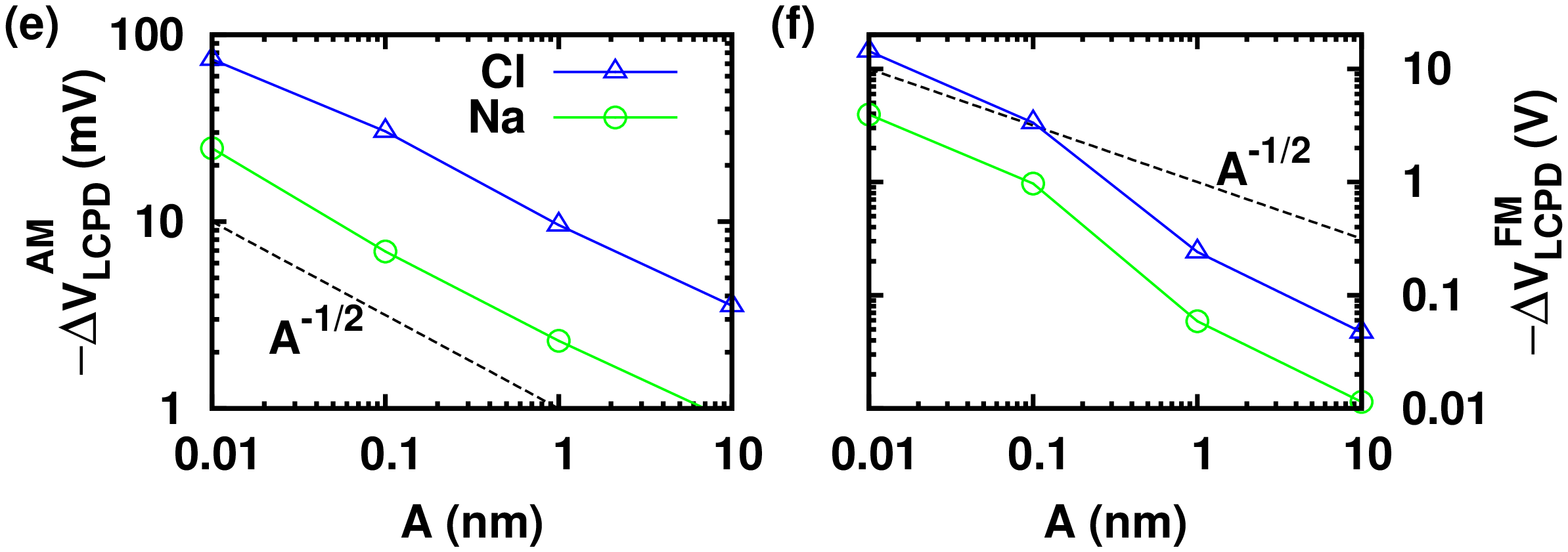}
\caption{\label{fig:ofA} (color online) Amplitude dependencies of
the cycle averages $\langle a \rangle$ and $\langle a'\rangle$ (a,b), $\langle C'\rangle$ and $\langle C''\rangle$ (c,d)
and of the resulting deviations $\Delta V_{LCPD}^{AM}$ and $\Delta V_{LCPD}^{FM}$
(e,f) at a closest tip apex separation of $d$=0.3~nm above Cl
and Na surface sites.}
\end{center}
\end{figure}

The same trends persist at all separations $d < 0.5$~nm where LCPD
contrast appears. $\langle a \rangle$ drops as $A^{-1/2}$, and $\langle a' \rangle$ drops as
$A^{-3/2}$ already beyond $A = 0.1$~nm, while $\langle C' \rangle$ varies only
little and $\langle C'' \rangle$ begins to drop somewhat slower than $A^{-1}$
only above $A = 1$~nm.  The resulting amplitude dependencies in
both modes reflect the different dependencies of the numerators
and denominators in Eqs.~(\ref{eqn:VL3}-\ref{eqn:VL4}).

\subsection{\label{sec:LCPD} Discussion and Experimental Limitations}
Expressions formally similar to 
Eqs.~(\ref{eqn:VL3}-\ref{eqn:VL4}) were obtained by Nony \emph{et
al}~\cite{Nony09a} who also noticed that $\langle a
\rangle$ and $A \langle a'  \rangle $ almost coincide when $A$
exceeds a few nanometers. However, their denominators came from a
\emph{short-range polarization contribution} $\propto V^2$ to the
\emph{microscopic force} $F_\mu$ rather than from the much larger
capacitive force $F_M$. 
This results in a comparable $\Delta V_{LCPD}$ for AM and FM modes
 if $A$ exceeds a few nanometers. 
However, by including the correct $F_M$
and taking into account the different amplitude dependencies of the denominators in
Eqs.~(\ref{eqn:VL3}-\ref{eqn:VL4}),
we conclude that 
the contrast should remain larger in the FM than in the AM mode for a
given closest approach distance $d$ and oscillation amplitude $A$.
This prediction is independent of the particular system
considered, but the mode-dependent signal to noise ratio 
must also be considered.  Thus Kawai \emph{et al.}~\cite{Kawai10}
 calculated the 
minimum detectable CPD as a function of $A$ and showed that it is smaller in the AM mode.
Taking into account the discussions of Figs.~\ref{fig:f} and
\ref{fig:CpCpp}, $ \langle C' \rangle$ would be larger if the
cantilever area is larger whereas  $\langle C'' \rangle$ would be
unaffected, whereas both quantities would be larger if the cone
angle is broader or if the sample is a metal rather than an
insulator, but $\langle C'' \rangle$ would be more strongly
affected. On the other hand $ \langle a \rangle$ and $ \langle a'
\rangle$ would be larger if the tip apex is charged~\cite{Nony09b}
rather than neutral, or if the sample is a semiconductor with a
reconstructed surface which exposes partially charged species like
Si(111) 7$\times$7 ~\cite{Sadewasser09,Kawai10}. From this point
of view the system studied here is especially challenging.
Furthermore, the contrast ratio slowly decreases if $A$ is
increased, e.g. by a factor which drops from about 100 to 10 for
oscillation amplitudes between 0.01 and 10 nm in our example.

For a meaningful comparison with NCAFM-KPFM measurements it is
important to take \emph{experimental limitations} into account. In
view of the long-range LCPD variations due to surface and bulk
inhomogeneities on real samples, one should compare computed
atomic-scale LCPD variations with the difference between the LCPD
measured at sub-nanometer separations $d$ in the middle of a flat
homogeneous island or terrace and the extrapolated long-range,
essentially site-independent LCPD. This procedure would also
suppress most of the long-range contributions to $ \langle a
\rangle$ and $ \langle a' \rangle$ which would arise in the case
of a charged or strongly polar tip~\cite{Barth10}. Moreover, the
comparison should be done with the same tip at constant $d$ (slow
distance control) because atomic-scale variations of $d$ at
constant $\Delta f_1(x,y,d)$ would induce such variations in the
LCPD even if the latter is site-independent but has a different
distance dependence as $\Delta f_1$.

For the distance controller to function properly, $\Delta f_1$
must be chosen on the branch where this frequency shift becomes
more negative if $d$ is decreased.  Furthermore, the maximum
restoring force $kA$ must be much larger than the maximum
tip-sample attraction~\cite{Giessibl97}. For measurements with
standard NCAFM cantilevers ($k\sim$~20-40~N/m) this criterion is
typically satisfied by using oscillation amplitudes $A >5$~nm,
and atomically resolved imaging is typically performed at
distances $d \sim$ 0.4-0.5~nm.  According to Fig.~\ref{fig:LCPD}
the LCPD contrast which is then predicted to be 20-100~mV in
the FM mode and a few mV in the AM mode 
approaches the experimental limits in both modes.
Even if the AM-KPFM
signal is enhanced by setting the modulation frequency at
the second flexural resonance of the
cantilever, 
the LCPD contrast predicted by our model would remain the same.
 This contrast would be stronger if the tip were charged. 
Unfortunately, available data showing
atomic-scale contrast on (001) surfaces of NaCl and KBr is
insufficient for a meaningful comparison between AM and FM KPFM.
However, LCPD maps obtained with sputter-cleaned Si tips and
similar measurement parameters on Si(111) 7$\times$7 surfaces show
that the contrast between Si adatoms and corner holes in the
FM-mode~\cite{Sadewasser09} is about ten times stronger than in
the AM-mode~\cite{Kawai10}. Moreover, data obtained from a direct
determination of the maximum of $\Delta f_1$ versus bias voltage
$V_b$ agreed well with those obtained by nulling the FM-KPFM
signal at the modulation frequency~\cite{Sadewasser09}.

The sizable LCPD contrast of several Volts predicted in the FM
mode for amplitudes $A <$ 0.1 nm should, however, be readily
observable when using a tuning fork instead of a cantilever. Owing
to the much higher stiffness $k \simeq$ 1800 N/m of this
deflection sensor, the above-mentioned criterion can be satisfied
with such amplitudes close to the ultrasmall
limit~\cite{Giessibl03}. Combined NCAFM-KPFM measurements using
such tuning forks with PtIr tips have only been done at low
temperature by the time-consuming direct method mentioned in the
Introduction.~\cite{Gross09a,Koenig09} Unfortunately, no FM-KPFM
measurements showing atomic-scale LCPD contrast on alkali halide
(001) surfaces have so far been reported.
\section{\label{sec:summary}Summary and outlook}
We proposed a general multiscale approach to compute electrostatic
forces responsible for atomic-scale contrast in KPFM performed
simultaneously with NCAFM.
 Although attention is focused on insulating samples and results
are presented for a silicon tip interacting with a NaCl(001)
sample, the approach is not restricted to particular sample or tip
materials.
The problem is split into two parts coupled in a remarkably simple
but novel fashion.
 First the electrostatic problem
 of the voltage-biased AFM probe (including the tip and the
cantilever) against the grounded sample, treated as macroscopic
perfect conductors or insulators, is solved by a finite-difference
method with controlled accuracy on a non-uniform mesh.
 The method is capable of treating complex geometries with widely
different dimensions, but is illustrated here for systems with
cylindrical symmetry.
 The solution yields the electric potential and field distributions and the
capacitance $C(s)$ of the system from which the electrostatic
force $F_M$ acting on the probe and its gradient are calculated as
functions of the macroscopic tip-sample separation $s$.
 By comparing results obtained with and without the cantilever, as
well as with the analytic solution for a tip approximated by a
conducting sphere in Appendix \ref{app:sphere}, the
\emph{contributions of the cantilever, the conical tip shank and
of its spherical end can be recognized. If the sample is a thick
insulator, all three affect the macroscopic force, whereas the
last two affect the force gradient even at sub-nanometer
separations relevant for atomic-scale contrast.}

Instead of the bias voltage $V_b$, the nearly uniform electric
field
 obtained in that range is then applied as an external field to the
microscopic part which can be treated by empirical atomistic or
first principles methods.  The \emph{ab initio} BigDFT wavelet
code employed here enables one to compute the short-range
bias-dependent force on the tip apex represented by a cluster with
an unprecedented accuracy of 1~pN per atom. For the
Si-nanotip-NaCl(001) system considered here, this microscopic
force $F_{\mu}$ is a linear function of the bias in the
investigated range $V_b - V_{CPD} = \pm$2 Volts. We argue that this is
a general result, except close to atomic-scale instabilities
caused by strong enough forces which could arise at very small
separations and/or very large effective biases.

Adding the macroscopic and microscopic bias-dependent forces,
expressions are obtained for the KPFM signals in the AM and the FM
modes.
 The atomic-scale deviation $\Delta V_{LCPD}$ of the local CPD from its
common value at large separations is the ratio of the derivatives
$a = dF_{\mu}/dV_b$ and $dC/ds$ averaged over the tip oscillation
amplitude with different weights in AM- and FM-KPFM, as described
by Eqs.~(\ref{eqn:VL3}-\ref{eqn:VL4}).
On the other hand, we explain the amplitude dependence of the atomic-scale LCPD contrast
in both modes and predict that for typical amplitudes used in measurements
with standard NCAFM cantilevers, this contrast
should be much stronger in the FM mode.  This is a consequence of
the contributions of the cantilever and the tip shank to the KPFM
signal in the AM mode, which are stronger on insulating samples.
The same conclusion has previously been reached in comparisons of
AM- and FM-KPFM measurements of long-range LCPD variations;
such variations are caused
by interactions of the biased probe with CPD inhomogeneities and
surface charges on scales of several nanometers and above on
conducting samples partly covered with ultrathin overlayers of
different materials~\cite{Glatzel03,Zerweck05}. However, the
strong mode-dependent influence of distant contributions to
$dC/ds$ on the atomic-scale LCPD contrast has, to our knowledge,
not been recognized because previous work on this topic assumed
that only the tip apex mattered at sub-nanometer separations.

Because $\Delta V_{LCPD}$ depends on measurement parameters, it is
desirable to \emph{extract the more fundamental quantity} $a =
dF_{\mu}/dV_b$ from combined KPFM measurements, just like the
microscopic force $F_{\mu}$ is extracted from NCAFM measurements
using, e.g. a widely accepted inversion algorithm~\cite{Sader04}
or one based on the direct inversion of the discretized version of
the first line of Eq.(\ref{eqn:VL4}) described in Appendix
\ref{app:averaging} by back-substitution~\cite{Pfeiffer04}.
 Since $\Delta V_{LCPD}$ is predicted to be stronger in FM-KPFM,
whereas its distance dependence is governed by the weighted
average $\langle a' \rangle$ modes, the most appealing way to
obtain $a(d)$ would be to extract $a'$ then integrate it from the
range where $\Delta V_{LCPD}$ vanishes down to the desired
separation $d$.  The averages $\langle a' \rangle$ and $\langle
C'' \rangle$ can be separately obtained from direct measurements
of the frequency shift $\Delta f_1$ as a function of bias
~\cite{Sadewasser09}, namely from the shift of the maximum and the
curvature of parabolic fits at several $(x, y, d)$ positions. 
The signal/noise ratio of those averages can be improved by using 
AC modulation and lock-in detection at the modulation frequency. 
 The averages could then be
determined from the zero intercept $V_{LCPD}^{FM}$ and the slope of the FM-KPFM signal
$\langle \Delta f_{\omega} \rangle$ versus DC bias.
An analogous procedure could be applied to determine $\langle a
\rangle$ and $\langle C' \rangle$ from the AM-KPFM signal $\langle
F_{\omega} \rangle$, then $a$ itself by inversion, using suitably
modified algorithms~\cite{Pfeiffer04,Sader10}.  
Because the AM-KPFM signal/ratio is much superior if the modulation frequency $f$ is at the second
cantilever resonance~\cite{Kawai10},
  $\Delta V_{LCPD}^{AM}$ could be determined more
accurately even if it is smaller than in  FM-KPFM. 
In any case,
note that the slope $a$ reflects variations of the electrostatic
potential outside the sample surface which are, however, locally
enhanced by the proximity of the tip apex. Since the latter is
in turn also polarized and deformed~\cite{Kawai11}, $a$ cannot
simply be described as the convolution of the unperturbed
electrostatic potential with a merely distance-dependent tip
point-spread function, as in macroscopic
electrostatics.~\cite{Strassburg05}

Complications due to averaging over the tip oscillation amplitude
are to a certain extent avoided with tuning fork deflection
sensors which enable direct measurements of $\langle \Delta
f_{\omega} \rangle$ vs. bias, using amplitudes approaching the
ultrasmall limit~\cite{Gross09a,Koenig09}.  Spectacular results
have thus been obtained on isolated molecules adsorbed on a thin
epitaxial NaCl(001) film by using tips with well-defined apex
species stable at low temperature~\cite{Gross09b}. Most recently,
$V_{LCPD}^{FM}$ contrast reflecting changes in the intramolecular
charge distribution has been observed upon a configurational
switch triggered by a judiciously applied pulse~\cite{Mohn12}. Our
results shown in Figs.~\ref{fig:LCPD}(e), \ref{fig:LCPD}(f) and \ref{fig:ofA}(d)
show that $\Delta V_{LCPD}^{FM}$ and $a'$ still have a significant
amplitude dependence between $A =$ 0.1 and 0.01 nm, so that
inversion is still necessary to obtain accurate results for
typical amplitudes used with tuning fork sensors.

Since such measurements use hard metal tips, while metal-coated
tips are also used in NCAFM and/or KPFM measurements with
cantilevers it would desirable to develop appropriate nanotip
models and to perform simulations like those described here.  In
particular, the recently fabricated sharp and stable 
W and Cr
coated silicon tips~\cite{Teobaldi11,Kinoshita11} and the stable
atomic-scale resolution achieved with Cr-coated cantilevers at
separations exceeding the usual range $d < 0.5$~nm merit further
attention.  Intentionally picked atoms or molecules at the apex
would be worth studying in a further step. Another class of
systems which merit further investigations involve silicon
nanotips with a picked-up cluster of foreign material, NaCl in
particular, which have so far been studied by DFT in the absence
of a sample~\cite{Amsler09} or represented by a cluster of the
same material as the sample using empirical interaction
potentials~\cite{Hoffmann04,Nony09b}.

Note finally that all macroscopic probe models, including ours,
 provide a better description of metallic or metal-coated tips 
than of real silicon tips.  Indeed, even if the native oxide
 is removed by sputtering, a silicon layer of few nanometers depleted
 of charge carriers still separates the tip surface from 
the highly doped conducting tip interior. 
  Although it was taken into account in previous treatments of
 KPFM of semiconductor devices~\cite{Sadewasser02}, 
 this depletion layer remains to be included when modelling Si tips,  
e.g. by allowing a smaller effective radius $R$ of the equipotential
 at the applied bias voltage and a larger effective separation $s$ from the sample surface.

\begin{acknowledgments}
This work has been supported by
the Swiss National Science Foundation (SNF) and
the Swiss National Center of Competence in Research (NCCR) on Nanoscale Science.
The CPU intensive computations were done at the Swiss National Supercomputing Center
(CSCS) in Manno.
\end{acknowledgments}

\appendix
\section{{Sign of the macroscopic electrostatic force} \label {app:sign}}
Using the virtual work method, the macroscopic electrostatic tip-sample interaction can be calculated
from the potential energy stored in the capacitor formed between the tip and the back-electrode.
The (real) force acting on the tip $F_z$, which is considered constant during a virtual arbitrary infinitesimal
 tip displacement  $\delta z$, performs a virtual work  $\delta w = F_z \cdot \delta z=-\delta U$, where
$U=U_c+U_b$ is the total energy of the system including contributions from both the capacitor and the biasing battery
which maintains a fixed potential difference $V$ between the both electrodes.
In response to this displacement,
the battery transfers a charge $\delta Q$ between the electrodes in order to keep their potential
difference fixed.
It costs a change of  $\delta U_b = -\delta Q \cdot V$ in the energy of the battery.
Whereas the energy of the capacitor changes by $\delta U_c = \frac{1}{2}\delta Q \cdot V$, which
implies $ U_b=-2U_c$, i.e.
\[\delta U=\delta U_c + \delta U_b=-\delta U_c.\]
The electrostatic force is therefore
\begin{align*}
F_z = -\frac{\delta U}{\delta z}= +\frac{\delta U_c}{\delta z}=  +\frac{1}{2}\frac{\delta C}{\delta z}V^2
\end{align*}
and is  always attractive because ${\delta C}/{\delta z}<0$.

\section{{Conducting sphere against a thick dielectric slab} \label {app:sphere}}
The force between a conducting sphere of radius $R$ at potential
$V$ facing a dielectric slab grounded on the bottom can be
calculated by means of the image charge method. For a
semi-infinite dielectric, we found that the solution is given by
remarkably simple generalization of the treatment in section 5.08
of Smythe's textbook~\cite{Smythe} for a semi-infinite conductor.
Details, further analytic results and useful approximations, which
are of general interest for scanning force microscopy, will be
presented elsewhere.~\cite{SadeghiAnalytic}
The potential $\Phi$ in the region between the sphere and the slab
is generated by a series of point charges $\{q_n,z_n\}$ inside the
sphere and their corresponding images $\{-\beta q_n,-z_n\}$ below
the surface of the dielectric, where
$\beta=(\epsilon-\epsilon_0)/(\epsilon+\epsilon_0)$, $\epsilon$
and $\epsilon_0$ being the permittivities of the dielectric and of
vacuum, respectively.
The first charge $q_1 = 4\pi \epsilon_0 V R$ is located at the
center of the sphere $z_1 = R+s$.  Physically, the image charges
represents the effect of the polarization induced at the surface
of the dielectric which causes a jump discontinuity in $E_z$.
Together with the other charges they ensure that
the sphere surface remains equipotential at $V$.

We find
\begin{eqnarray} \label{eqn:qn}
q_n &=& q_1 \sinh \alpha 
\Big(\frac{\beta^{n-1}}{\sinh n\alpha} \Big), 
\end{eqnarray}
where $\cosh \alpha=z_1/R$ and
\begin{eqnarray} \label{eqn:zn}
z_n &=& R\sinh\alpha\coth n\alpha,
\end{eqnarray}
as in Smythe's treatment ($\beta = 1$).  Except at the contact point
($s$ = 0), the charges $q_n$ decay 
exponentially fast
towards zero and this solution provides convenient expressions for
the capacitance $C(s) = q_{sph}/V$, where
\begin{eqnarray} \label{eqn:qeff}
q_{sph}=\sum_{n=1}^\infty q_n= 4\pi \epsilon_0 R V \sinh \alpha \times
   \sum_{n=1}^\infty  \frac{\beta^{n-1}}{\sinh n\alpha },
\end{eqnarray}
and the z-component of the force $(dC/ds)V^2/2$
\begin{eqnarray} \label{eqn:Fcompact}
F_M=
2\pi \epsilon_0 V^2
  \sum_{n=2}^\infty  \frac{\beta^{n-1}}{\sinh n\alpha } \big(\coth \alpha -n\coth n\alpha \big),
\end{eqnarray}
the force gradient $dF_M/ds$,
and the electric field at $\rho=0, z=0$
\begin{eqnarray} \label{eqn:E0compact}
E_z= \frac{V}{R}
 \frac{1+\beta}{\sinh \alpha}
  \sum_{n=1}^\infty  \frac{\beta^{n-1}\sinh n\alpha}{\cosh^2 n\alpha }.
\end{eqnarray}
These series have been used to evaluate the solid lines in Figs.~\ref{fig:f} and \ref{fig:Eofs}(b)
The truncation error can be reduced 
by generalizing a trick proposed for the conducting sphere-plane
problem~\cite{Kantorovich00}.
If a series is truncated at some $n = k$, the remainder
can be summed up analytically if one assumes $z_{n>k} \simeq
z_\infty =R \sinh\alpha$.
Thus, because $q_{n+1}/q_{n>k}\simeq \beta R/(z_1+z_\infty)$,
\[ q_\infty \equiv \sum_{n=k+1}^\infty q_n \simeq \frac{q_{k+1}}{1-{\beta R}/{(z_1+z_\infty)}}=
  \frac{q_{k+1}}{1-\beta e^{-\alpha}}.  \]
Adding the correction to the first 5 terms, $q_{sph}$ is obtained with an accuracy of $10^{-5}$
 for $\epsilon/\epsilon_0=5.9, s/R=0.1$.

For an ideal conductor ($\beta=1$) these expressions reduce to
those of Smythe~\cite{Smythe} and diverge in the limit $s \to 0$
(i.e. $\alpha \to 0)$. In our case $\beta<1$, and in the same
limit the resulting series converge and can in fact be summed explicitly.~\cite{SadeghiAnalytic}
  The result for $F(s=0)$ was given without proof in Eq. (2) of
Ref.\cite{Gomez01b}. 
For NaCl ($\epsilon/\epsilon_0=5.9$, $\beta=0.71$) one obtains
limiting values of $C/\pi \epsilon_0 R=6.98$, $F/\pi \epsilon_0
V^2=-6.77$ (i.e. $F=-0.188\text{ nN/V}^2$ independent of sphere
radius), $F'/\pi \epsilon_0V^2R^{-1}=188.7$ and $E/VR^{-1}=20.4$.
%
%
In the case of a dielectric slab of finite thickness $t$, mirror
images of all previously mentioned charges with respect to the
grounded back-electrode must also be considered. They ensure that
the field lines inside the dielectric become perpendicular to the
grounded back-electrode instead of spreading radially.
If $z_1 =R+s \ll t$, further images charges induced by those
mirror charges can be neglected to order $\mathcal O((z_1/t)^2)$.
Moreover,
 the electric force exerted by the mirror charges on the biased
sphere can then be approximated as the Coulomb force between
$q_{sph}$ at its center and a lumped mirror charge
$-(1-\beta)q_{sph}$,
$2(z_1+t)$ away, i.e.
\[
F_{add} \simeq \frac {-(1-\beta)} {4\pi \epsilon_0} \frac {q_{sph}^2}{(2t)^2}=
\pi \epsilon_0V^2 \frac{2\epsilon_0}{\epsilon +\epsilon_0} \Big(\frac{R}{t}\Big)^2 \Big(\frac{q_{sph}}{q_1}\Big)^2
.\]
For relevant values $s<R\sim$10 nm and $t \sim$ 1 mm, this
correction is below
$(R/t)^2/6.8  \sim 10^{-11}$
times the force given by
Eq.~(\ref{eqn:Fcompact}),
 i.e. negligible in practice.
  A similar expression of comparable magnitude was proposed in
 Ref.~\cite{Nony09a}, but was erroneously assumed to represent $F_M$.

\section{Discretized integrals for finite tip oscillation amplitudes \label{app:averaging}}
Assuming that $N+1$ equispaced data points $\{z_i\}$ 
are sufficiently close together such that $g(z)$ remains almost constant within an interval length $\delta={2A}/{N}$,
 the integration in Eq.~(\ref{eqn:<g>}) 
can be approximated by a  finite sum
\begin{eqnarray}
\langle g(z) \rangle 
\simeq \frac{1}{\pi}  \sum_{i=0}^{N}  W_i g_i \nonumber
\end{eqnarray}
where
 $g_i \equiv g(z_i)$ is either $g_i= a(z_i)$ or $g_i=C(z_i+h)$;
Since $W(\zeta)={1}/{\sqrt{A^2-\zeta^2}}$ we obtain
\begin{eqnarray}
W_i=\int_{\zeta_i^-}^{\zeta_i^+} W(\zeta) d\zeta =
 \arcsin( \frac{\zeta_i^{+}}{A} )
-\arcsin( \frac{\zeta_i^{-}}{A} ) \nonumber
\end{eqnarray}
where $\zeta_i^{\pm}=(i\pm\frac{1}{2}) \delta - A$
 are the midpoints between $\zeta_i$ and $\zeta_{i \pm 1}$.
 Taking into account the rapid variation of $W(\zeta)$ near the
integration limits
 defined as $\zeta_0^-=-A$ and $ \zeta_N^+=A$, the square root
singularities of $W(\zeta)$ at those turning points are
approximately included with this modified trapezoid integration
method. Sufficiently far from those points $W_i\simeq
W(\zeta_i)\delta $ so that the standard trapezoid approximation
 is recovered. The analogous approximation for Eq.(\ref{eqn:<gp>}) namely
\begin{eqnarray}
\langle g'(z) \rangle 
\simeq \frac{1}{\pi}  \sum_{i=0}^{N}  W^*_i g_i \nonumber
\end{eqnarray}
involves~\cite{Pfeiffer04}
\begin{eqnarray}
W^*_i&=&\frac{1}{A} \int_{\zeta_i^-}^{\zeta_i^+} \zeta W(\zeta) d\zeta  =
\sqrt{1-\Big(\frac{\zeta_i^-}{A}\Big)^2}- \sqrt{1-\Big(\frac{\zeta_i^+}{A}\Big)^2} \nonumber.
\end{eqnarray}
Note that in the $A \to 0$ limit only the data points at the two
limits are taken into account. Indeed, if  $N=1$, $A=\delta/2$ and
$W_0=W_1$, hence $\langle g \rangle=(g_0+g_N)/2$,
 and $W_0^*=-W_1^*$, hence $\langle g' \rangle=(g_N-g_0)/2A$,
so that Eqs.~(\ref{eqn:VL3},\ref{eqn:VL4})
consistently approximate the corresponding zero-amplitude
equations, Eqs. (\ref{eqn:LCPDAM},\ref{eqn:LCPDFM}).
Similarly, if $N$=2, $A=\delta$ and one obtains $W_0=W_2,W_1=0$
and $W_0^*=-W_2^*,W^*_1=0$ and
Eqs.~(\ref{eqn:LCPDAM},\ref{eqn:LCPDFM}) are again recovered.

\nocite{*}

\bibliography{lcpd}
\end{document}